\documentclass[%
 twocolumn, 
 longbibliography,
 aps,
 prb,
 floatfix,
 superscriptaddress
]{revtex4-2} 

\usepackage[english]{babel}
\usepackage{lmodern}

\usepackage[dvipsnames]{xcolor}

\usepackage{amsmath, amssymb}
\usepackage{graphicx}
\usepackage[utf8]{inputenc}
\usepackage{dsfont}
\usepackage[T1]{fontenc}
\usepackage{braket}
\usepackage{mathtools}
\usepackage{physics}
\usepackage{tikz-network}
\usepackage{pifont}
\usepackage{CJKutf8}
\usepackage{bm}
\usepackage{float}
\usepackage[colorlinks=true,breaklinks=true,linkcolor=blue,anchorcolor=red,citecolor=blue,urlcolor=blue]{hyperref}
\usepackage{bbm}

\usepackage{dcolumn}

\usepackage{lipsum}
\usepackage{comment}
\newcommand{\cmark}{\ding{51}}%
\usepackage{tabularx}

\usepackage{multirow}
\newcommand{\vect}[1]{\boldsymbol{#1}}

\def\cH{\mathcal{H}}

\def\cO{\mathcal{O}}

\renewcommand{\d}{\mathrm{d}}

\begin{document}
\begin{CJK*}{UTF8}{gbsn}
\title{Triangular lattice models of the Kalmeyer-Laughlin spin liquid from coupled wires}
\author{Tingyu Gao (高廷宇)}
\affiliation{Department of Condensed Matter Physics, Weizmann Institute of Science, Rehovot 7610001, Israel}
\affiliation{Institute for Theoretical Physics, University of Cologne, D-50937 Köln, Germany}

\author{Niklas Tausendpfund}
\affiliation{Institute for Theoretical Physics, University of Cologne, D-50937 Köln, Germany}
\affiliation{Forschungszentrum Jülich GmbH, Institute of Quantum Control, Peter Grünberg Institut (PGI-8), 52425 Jülich, Germany}

\author{Erik L. Weerda}
\affiliation{Institute for Theoretical Physics, University of Cologne, D-50937 Köln, Germany}

\author{Jan Naumann}
\affiliation{Freie Universität Berlin, Dahlem Center for Complex Quantum Systems, D-14195 Berlin, Germany}

\author{Matteo Rizzi}
\affiliation{Institute for Theoretical Physics, University of Cologne, D-50937 Köln, Germany}
\affiliation{Forschungszentrum Jülich GmbH, Institute of Quantum Control, Peter Grünberg Institut (PGI-8), 52425 Jülich, Germany}

\author{David F. Mross}
\email{david.mross@weizmann.ac.il}
\affiliation{Department of Condensed Matter Physics, Weizmann Institute of Science, Rehovot 7610001, Israel}

\begin{abstract} 
Chiral spin liquids (CSLs) are exotic phases of interacting spins in two dimensions, characterized by long-range entanglement and fractional excitations. We construct a local Hamiltonian on the triangular lattice that stabilizes the Kalmeyer-Laughlin CSL without requiring fine-tuning. Our approach employs coupled-wire constructions and introduces a lattice duality to construct a solvable chiral sliding Luttinger liquid, which is driven toward the CSL phase by generic perturbations. By combining symmetry analysis and bosonization, we make sharp predictions for the ground states on quasi-one-dimensional cylinders and tori, which exhibit a fourfold periodicity in the circumference. Extensive tensor network simulations demonstrating ground-state degeneracies, fractional quasiparticles, nonvanishing long-range order parameters, and entanglement signatures confirm the emergence of the CSL in the lattice Hamiltonian.
\end{abstract}
\date{\today}
\maketitle
\end{CJK*}

\section{Introduction} \label{introduction}

Interacting spins on a lattice typically realize ground states that break internal symmetries, such as spin rotations or lattice symmetries.
Competition between interactions that favor different ordering patterns can result in enhanced quantum fluctuations that instead promote quantum spin liquids; see Refs.~\cite{balents2010spin, Savary16QSLReview, zhou2017, wen2019, broholm2020sci} for recent reviews. 
These strongly correlated states are characterized by topological order instead of symmetry-breaking and exhibit long-range entanglement. 

The Kalmeyer-Laughlin chiral spin liquid (CSL) \cite{Kalmeyer87CSL} is a paradigmatic example of this class of phases. 
It shares many qualitative features with fractional quantum Hall (FQH) states, particularly, the $\nu = 1/2$ Laughlin state of bosons \cite{Laughlin83prl}. 
Both phases exhibit a topological degeneracy on a torus, chiral edge states at an open boundary, and fractional quasiparticles that exhibit anyonic exchange statistics. 
Microscopically, the FQH and CSL differ significantly regarding the relevant length scales and their competition with other phases.
Quantum Hall states are most easily realized in continuous systems or at low densities where the interparticle spacing is the only relevant length scale. By contrast, quantum spin liquids are governed by lattice scale effects and compete energetically with conventional, symmetry-broken phases. 
Consequently, finding generic lattice Hamiltonians whose ground states realize spin liquids represents a significant challenge. 

For the Kalmeyer-Laughlin CSL, Refs.~\cite{Schroeter07PRL, Thomale09PRB, Nielsen12PRL} addressed this challenge by constructing a Hamiltonian that explicitly projects onto a specific many-body wave function representing this phase. 
A less direct but more widely applicable route involves ``coupled wire'' constructions. This approach was pioneered for FQH states in Ref.~\cite{Kane02prl} and adapted for CSLs in Refs.~\cite{Meng15prb, Gorohovsky15prb}.
It begins by analytically solving decoupled one-dimensional subsystems (wires) using field-theory techniques.
Carefully selected couplings between the wires are then introduced into the effective field theory to form a two-dimensional phase.

In this article, we combine the coupled-wire approach with tensor-network methods~\cite{White92DMRG, Zauner18PRB} to introduce a local Hamiltonian (up to four-spin terms) for the CSL on the triangular lattice and confirm its ground state. 
The crucial step in our approach is the construction of the gapless sliding Luttinger liquids (SLL) with the CSL as the strongest instability. The SLL fixed point is characterized by a separate $U(1)$ symmetry for each wire in the array. Such symmetry requirement is, however, not a sufficient condition for having an SLL. In general, even interchain interactions that preserve these $U(1)$ symmetries induce instabilities to gapped phases that preempt the formation of the SLL. 
Here, we employ a duality relation to identify nonperturbative interwire couplings that realize the SLL fixed point with certainty. Moreover, we can analytically control the universal parameters of the SLL through the coupling constant of local lattice operators.

Our analysis of the coupled-wire model reveals how the CSL manifests itself in quasi-one-dimensional systems. 
In particular, we analyze the quasi-one-dimensional avatars of the topologically degenerate sectors expected in two dimensions. 
On cylinders formed by $N$ wires, we find a surprising fourfold periodicity. 
When $N$ is not a multiple of four, the one-dimensional ground state exhibits weak symmetry breaking \cite{wang2013} with an order parameter that vanishes exponentially with $N$.
When $N$ is a multiple of four, the one-dimensional ground state is unique and symmetric.
In this case, the different topological sectors of the two-dimensional phase correspond to different symmetry-protected topological (SPT) states in one dimension, which become degenerate as $N \rightarrow \infty$.

We tested the detailed predictions based on the coupled wire analysis using extensive numerical simulations. The ground states obtained from density matrix renormalization group (DMRG)~\cite{White92DMRG} and variational uniform matrix product state (VUMPS)~\cite{Zauner18PRB} techniques on narrow-width (finite and infinite) cylinders confirm the expected signatures of quasi-one-dimensional CSL avatars. 
In particular, we investigated ground-state degeneracies, symmetry breaking, string order parameters, entanglement spectra, and spin pumping. 
Additionally, we employed infinite projected-entangled pair states (iPEPS)~\cite{verstraete2004PEPS, Jordan2008iPEPS} to show that, by use of variational optimization~\cite{Corboz16Vari, Liao19ADiPEPS, Weerda24VariPEPS}, we can find a chiral spin liquid state in the proposed model and characterize it via its entanglement spectrum. 

This article is organized as follows: In Sec.~\ref{Coupled-wire}, we briefly review the essential aspects of the Kalmeyer-Laughlin CSL and how to construct it from coupled wires. In particular, we discuss the relevant symmetries and identify competing phases, such as valence bond solids and antiferromagnets. In Sec.~\ref{SLL}, we engineer a lattice model for the SLL and derive its field theory using a combination of duality and one-dimensional bosonization. In Sec.~\ref{CSL1D}, we analyze CSL signatures in the quasi-one-dimensional limit. Section \ref{Numerics} presents our numerical results, which support a CSL ground state. We conclude in Sec.~\ref{Discussion} with a brief discussion of our main results and possible generalization to more exotic spin liquids.

\section{Chiral spin liquid and coupled-wire construction} \label{Coupled-wire}

Spin systems in two or higher dimensions usually settle into conventional, symmetry-breaking phases, such as valence bond solids, antiferromagnets, or ferromagnets.
However, it is now well established that symmetry-preserving but topologically ordered phases are also possible~\cite{balents2010spin, Savary16QSLReview}. 
Such states may form when frustration between different interactions enhances quantum fluctuations and suppresses conventional orders. 
A prominent example of frustration occurs when spins interact anti-ferromagnetically on the triangular lattice.
The three spins at the corners of each triangular plaquette cannot be simultaneously antiparallel to both of their neighbors.
Consequently, the resonating valence bond state---a $\mathbb{Z}_2$ quantum spin liquid in modern parlance---was proposed as a possible ground state~\cite{Savary16QSLReview}.

Seminal work by Kalmeyer and Laughlin revealed an intimate connection between frustrated spin models and quantum Hall systems~\cite{Kalmeyer87CSL}. 
Spin-exchange interactions are mathematically equivalent to hopping processes and interactions of hard-core bosons.
The frustration translates into a non-zero background flux for these bosons.
On the triangular lattice, this flux is $\pi$ per triangular plaquette.
Summing these fluxes yields a total phase of $2\pi$ per unit cell.
In the zero magnetization sector, where the bosons are at half-filling of the lattice, the Landau-level filling factor is $\nu=1/2$, allowing the bosons to form a Laughlin state.

The ground state of the nearest-neighbor Heisenberg model on the triangular lattice was later shown to be in a N\'eel state~\cite{Capriotti99_PRL, White07PRL}.
Still, the Kalmeyer-Laughlin CSL is an energetically competitive state that could be stabilized by additional next-nearest-neighbor or multispin interactions. 
References \cite{Schroeter07PRL, Thomale09PRB} built on the quantum Hall connection to construct exact parent Hamiltonians for the Kalmeyer-Laughlin wave function. Soon after, Ref.~\cite{Nielsen12PRL} proposed another exact model by studying a modified wave function in light of conformal field theory. In particular, this model can be truncated into a short-range Hamiltonian consisting of two- and three-spin interactions that realize the same phase \cite{Nielsen13ncomms}.

Here, we pursue a different strategy based on a perspective that originates in fractional quantum Hall studies.
Coupled-wire models have become a popular theoretical tool for analytically capturing topological states of electrons, bosons, or spins~\cite{Kane02prl, Teo14PRB, Meng15prb, Gorohovsky15prb}.
They do not yield exact microscopic Hamiltonians like the ones in Refs.~\cite{Schroeter07PRL, Thomale09PRB, Nielsen12PRL}. However, in return, they offer much broader applicability.

\subsection{Coupled-wire construction of quantum Hall states and chiral spin liquids}

\begin{figure}[tb]
    \includegraphics[width=\columnwidth]{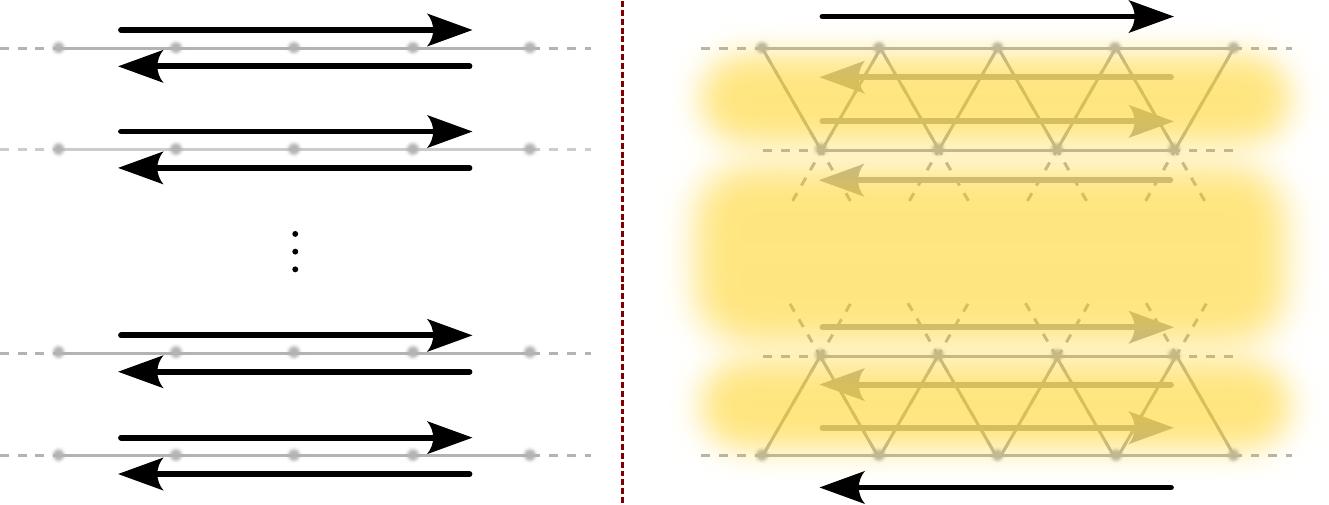}
    \caption{Schematic figure for the coupled-wire construction. Gapping out counterpropagating modes on the nearest chains gives an inert bulk. The unpaired modes remain gapless at the boundary.}
    \label{fig.Wires}
\end{figure}

The main idea behind the coupled-wire description of chiral topological states is shown in Fig.~\ref{fig.Wires}.
Its starting point is an array of electronic quantum wires or spin chains in the gapless phase, governed by the Hamiltonian density
\begin{equation} \label{eqn.LLxxz}
\cH_{\text{LL}}\left[\varphi, \theta\right] = \frac{u}{2\pi}  \, \left[\frac{1}{K}\left(\nabla \theta \right)^2 + K \left(\nabla \varphi \right)^2\right]
\end{equation}
of each wire. It is written in terms of two conjugate bosons, which obey the commutation relation $\comm{\partial_x \varphi(x)}{\theta(x')} = i \pi \delta(x-x')$. 
The two fields are equivalent to a pair of left- and right-moving modes $\phi^{n}_{L/R} =\varphi \pm n \theta$, which decouple when the Luttinger parameter $K=1/n$.

Quantum Hall states or chiral spin liquids arise when left- and right-moving modes on neighboring wires gap out in pairs, leaving unpaired chiral modes at the boundary. 
To achieve such a topological state, the array must be coupled via a relevant perturbation in the renormalization group (RG) sense. 
A low-energy operator that can induce a chiral state is given by
\begin{align} \label{eqn.ham.chir}
    {\cal{O}}_\text{chiral}^n = \cos[\phi^{n}_{L,y}-\phi^{n}_{R,y+1}]~,
\end{align}
where $y$ labels the wires, and $n$ must be even (odd) for spins (fermions). In the case of continuous translations in $x$, this operator is only kinematically allowed to appear in the Hamiltonian in a magnetic field corresponding to the filling factor $\nu=1/n$. The scaling dimension of $ {\cal{O}}_\text{chiral}^n$ according to ${\cal H}_0 = \sum_y {\cal H}_\text{LL}[\varphi_y, \theta_y]$ is
\begin{align}
\Delta_n(K)=\frac{1}{2K}+\frac{n^2}{2}K~,
\end{align}
whose minimum is $\min_K \Delta_n(K) = n$ (for $K=1/n$). 
The decoupled array is invariant under a rescaling of time and the wire direction $x$ but not the transverse direction $y$. 
Consequently, an operator is relevant if the scaling dimension $\Delta < 2$. For spin systems, $n$ must be even; the CSL with the smallest $\Delta$ thus occurs for $n=2$ and will be our focus. In the decoupled array, it can, at best, be marginal (at $K=1/2$).

Gapped phases accessed via marginally relevant perturbations would exhibit small energy gaps and large correlation lengths, which are obstacles to efficient numerical studies. 
Moreover, competing phases driven by more relevant operators can preempt the flow toward the CSL and result in a different ground state.
The possible competing phases are dictated by the lattice structure, and we focus on the triangular lattice for its built-in frustration.

Before closing, we note that for a chiral state without fractionalization, the integer quantum Hall state of bosons at $\nu=2$ \cite{Lu12PRB, Geraedts2013aop, Senthil13PRL, Furukawa13PRL}, the corresponding interwire coupling can be tuned relevant by intrawire couplings alone.
Specifically, the interwire coupling $\cos(\varphi_{y+1} - 2 \theta_{y} -\varphi_{y-1})$ has scaling dimension $\Delta_\text{BIQH}(K) = \frac{1}{2K} +K$, with the minimum $\Delta_\text{BIQH}(1/\sqrt{2})=\sqrt{2}$. 
Under this condition, the connection between coupled-wire analysis and lattice models is relatively straightforward, as discussed in Refs.~\cite{Fuji15PRB, He15PRL}.

\begin{figure}[tb]
    \includegraphics[width=\columnwidth]{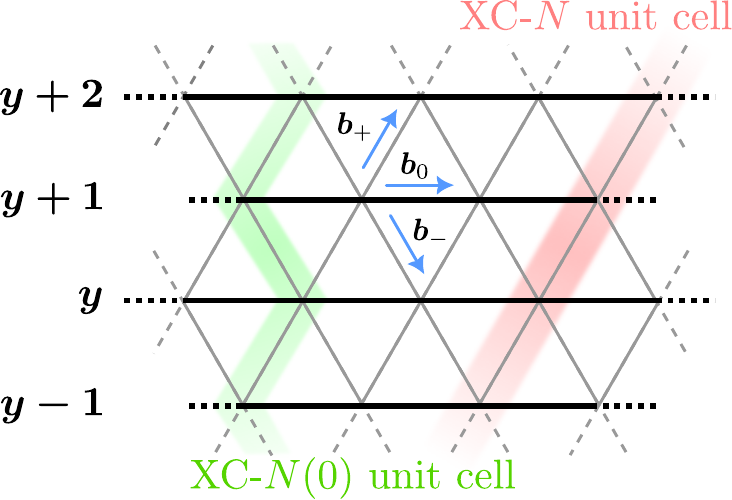}
    \caption{Triangular lattice is defined by the unit vectors $\bm{b}_0$, $\bm{b}_+$, and $\bm{b}_-$ connecting neighboring sites. On a cylinder (shown here for $N=4$), different transverse boundary conditions are possible. The XC-$N$ boundary condition is equivalent to choosing the red unit cell. The XC-$N(0)$ boundary condition uses the green zig-zag unit cell, which is only possible when $N$ is even.
    }
    \label{Lattice}
\end{figure}

\subsection{Symmetries and competing phases on a triangular array}

We consider a triangular array defined by fundamental lattice vectors $\bm{b}_0 = \hat{x}$ and $\bm{b}_{\pm}= \frac{1}{2} \hat{x} \pm \frac{\sqrt{3}}{2} \hat{y}$, as illustrated in Fig.~\ref{Lattice}. 
The six-fold rotation symmetry is broken in the anisotropic limit of coupled wires, and we retain only translation and parity (mirror) symmetries.
To realize the CSL, we break the time-reversal symmetry $({\cal T})$ and the mirror symmetry $({\cal P})$ while preserving their product.
We also relax the SU(2) spin rotation symmetry and require only invariance under arbitrary rotations around $S^z$ and under a $\pi$ rotation around $S^x$. The latter symmetry implies an unpolarized state with $\langle S^z\rangle =0$, equivalent to the half-filling of hard-core bosons.
The discrete symmetries are summarized in Tab.~\ref{symmetries}.

\setlength{\extrarowheight}{3pt}
\setlength\tabcolsep{5pt}
\begin{center}
\begin{table}[tb]
\caption{\label{symmetries} Transformations of the lattice spins $\vec S$ and the bosonized variables $\theta,\varphi$ under discrete symmetries. 
The parity symmetry ${\cal P}$ involves a mirror centered on \textit{sites} of even-numbered wires (where $x$ are integer) and on \textit{links} of odd-numbered wires (where $x$ are half-integer). Half-integer $x$ in the oscillatory terms of Eqs.~\eqref{eqn.Sz} and \eqref{eqn.Sp} imply the different transformations of $\theta,\varphi$ listed below.
}

\begin{tabularx}{\linewidth}{c|ccccc}
\hline\hline
 & $\vec S$&$(x,y)$& $\theta$ & $\varphi$&\\
 \hline
$T_x$&$\vec S$ &$(x+1,y)$ &$\theta + \frac{\pi}{2}$ & $\varphi  + \pi$ &\\
$\Pi_x$  &$(S^x,-S^{y,z})$ &$(x,y)$ &$-\theta + \frac{\pi}{2}$ & $-\varphi$& \\
${\cal T}$ &$-\vec S$ &$(x,y)$ &$-\theta + \frac{\pi}{2}$ & $\varphi  + \pi$& \\
\multirow{2}{*}{${\cal P}$}  &\multirow{2}{*}{$\vec S$} &\multirow{2}{*}{$(-x,y)$} &$-\theta$ & $\varphi $ &(even $y$)\\
  && &$-\theta + \frac{\pi}{2}$ & $\varphi + \pi$ &(odd $y$)\\
\hline \hline
\end{tabularx}
\end{table}
\end{center}

The spin Hamiltonian 
\begin{align}\label{eqn.hxxz}
H_\parallel &= \frac{1}{2} \sum_{\bm{r}} \left( S^+_{\bm{r}} S^-_{\bm{r} + \bm{b}_0} + \text{H.c.} \right) + J^z_\parallel \sum_{\bm{r}} S^z_{\bm{r}} S^z_{\bm{r} + \bm{b}_0}~
\end{align}
describes an array of decoupled chains, each of which realizes Eq.~\eqref{eqn.LLxxz} for $J^z_\parallel \in (-1,1)$. 
The parameters $u,K$ are known from the Bethe ansatz solution to be~\cite{Giamarch04book}
\begin{align}
K=\frac{\pi}{2}\frac{1}{ \cos ^{-1}(-J^z_\parallel)}~, \quad 
u=\frac{\pi}{2}\frac{  \sqrt{1-(J^z_\parallel)^2}}{ \cos ^{-1}(J^z_\parallel)}
  \label{eqn.ba}  ~.
\end{align}
The microscopic spin operators at $\vect r = (x,y)$ are expressed in terms of the long-wavelength variables $\varphi,\theta$ in Eq.~\eqref{eqn.LLxxz} as
\begin{subequations}
\begin{align}
    S^z_{\bm{r} } &\sim -\frac{1}{\pi}\partial_x \theta_y(x) + c_1 (-1)^x \cos{\left[2 \theta_y(x)\right]}~, \label{eqn.Sz}\\
    S^+_{\bm{r}} &\sim c_2 e^{-i \varphi_y(x)} \cos{\left[2 \theta_y(x)\right]} + c_3 (-1)^x e^{ -i \varphi_y(x)}~, \label{eqn.Sp}
\end{align}
\end{subequations}
with nonuniversal constants $c_{1,2,3}$. From these expressions, we infer the action of different discrete symmetries on the bosonized variables, listed in Table~\ref{symmetries}.

It is instructive to use this dictionary to determine how the geometric frustration of the triangular lattice manifests itself in the bosonized variables \cite{Nersesyan98prl}.
In particular, we consider the interwire spin exchange between neighboring spins, i.e., 
\begin{subequations}
\begin{align}
\label{eqn.hperp}
   H_\perp &= \frac{J_\perp}{2} \sum_{i = \pm} \sum_{\bm{r}} \left(S^+_{\bm{r}} S^-_{\bm{r} + \bm{b}_{i}} + \text{H.c.}\right)\\
    & = \frac{J_\perp}{2} \sum_{\bm{r}} \left[ \left(S^+_{\bm{r}} + S^+_{\bm{r} + \bm{b}_0}\right) S^-_{\bm{r} +\bm{b}_+} + \text{H.c.}\right]~.
\end{align}
\end{subequations}
When inserting Eq.~\eqref{eqn.Sp} into the second line, the oscillatory contributions from $S^+_{\bm{r}}$ and $ S^+_{\bm{r} + \bm{b}_0}$ cancel. We thus obtain
\begin{equation}\label{eqn.spinexchange}
    {\cal H}_\perp\sim g \sum_y \, [{\cal O}_\text{chiral}+{\cal O}_\text{antichiral}] + \cdots ~,
\end{equation}
with ${\cal{O}}_\text{chiral}$ given by Eq.~\eqref{eqn.ham.chir} at $n=2$ and ${\cal{O}}_\text{antichiral}$ its time-reversed counterpart.
In particular, the coupling $\cos{(\varphi_y - \varphi_{y+1})}$, which promotes $XY$ magnetic order, is absent, reflecting the intrinsic frustration of the triangular lattice. By contrast, that cosine is generically present on a square lattice, which is not frustrated and does not exhibit a similar cancellation. More rigorously, the absence of this coupling follows from the combined symmetry of parity $x \rightarrow -x$ and time reversal defined in Table \ref{symmetries}. However, the cancellation invoked above is incomplete, and the ellipsis in Eq.~\eqref{eqn.spinexchange} includes \cite{Nersesyan98prl}
\begin{equation}
   {\cal{O}}_\text{spiral} =\left(\grad \varphi_y + \grad \varphi_{y+1}\right) \sin{\left(\varphi_y - \varphi_{y+1}\right)}~.
\end{equation}
This operator is marginal for $K=1/2$ and thus directly competes against the CSL.

\begin{figure}[tb]
    \includegraphics[width=\columnwidth]{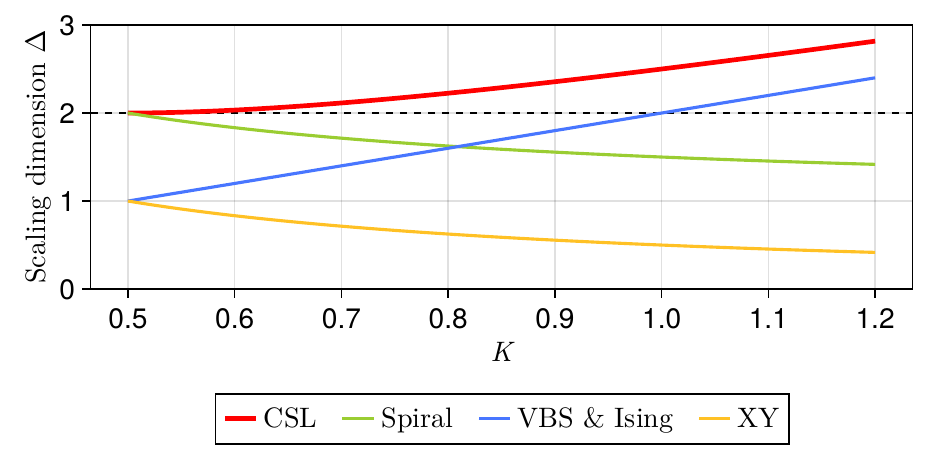}
    \caption{Scaling dimensions at the decoupled Luttinger liquid fixed point. The operator $\cO_{\text{chiral}}$ leading to the CSL has the largest scaling dimension. In particular, this value is greater or equal to $2$ for any $K$, meaning it is at most marginal. In contrast, the scaling dimensions of the other operators can be smaller than $2$, in which case they are relevant.}
    \label{fig.XXZscalings}
\end{figure}

Even more dangerous to the CSL are interactions between second-neighbor wires, which are free of frustration and are generated at order $J_\perp^2$. The most important ones are 
\begin{subequations}
\begin{align}\label{Neel}
&{\cal{O}}_{XY} = \cos{\left(\varphi_y - \varphi_{y+2}\right)}~,\\
&{\cal{O}}_\text{Ising} = \cos{\left(2 \theta_y\right)} \cos{\left(2\theta_{y+2}\right)}~,\\
&{\cal{O}}_\text{VBS} =  \sin{\left(2 \theta_y\right)} \sin{\left(2\theta_{y+2}\right)}~.
\end{align}
\end{subequations}
When either of these operators determines the low-energy physics, they realize independently ordered states on all even and all odd wires. 
For example, the operator ${\cO}_{XY}$ can lead to a striped phase with antiferromagnetic order along $x$ and ferromagnetic order along $y$ between every two chains. 
One approach to quantify the competition is to compare the scaling dimensions. For the competitors above, we find
\begin{equation}
\label{eq.dimensions}
\begin{aligned}&\Delta_{XY} = \frac{1}{2K}~,  \qquad \quad
&&\Delta_{\text{chiral}} = \frac{1}{2K} + 2K~, \\
&\Delta_{\text{Ising}} = \Delta_{\text{VBS}} = 2K~, \quad 
&&\Delta_{\text{spiral}} = \frac{1}{2K}+1 ~.
\end{aligned}\end{equation}
In particular, at the value $K=1/2$, where the perturbation ${\cal{O}}_\text{chiral}$ is (marginally) relevant, two strongly relevant perturbations drive the system into competing phases. Figure \ref{fig.XXZscalings} shows all four dimensions for $K$ in the interval $(0.5,1.2)$.
Therefore, to achieve the desired $\nu=1/2$ CSL, we resort to a different starting point, which we detail in the next section.

\section{Chiral sliding Luttinger liquid} \label{SLL}

To promote the CSL, we would like to access it via a relevant perturbation and simultaneously suppress the competing phases. 
Both can be achieved by generalizing the decoupled array to a coupled one that realizes a SLL fixed point at low energy~\cite{Kane02prl}. 
The main challenge to constructing a suitable SLL is that generic interwire couplings often drive the wire array into one of the competing symmetry-broken phases.
We overcome this difficulty first by designing interwire couplings that maintain integrability, which we prove below using a duality relation. 
The key ingredient for the SLL is the spin rotation operator 
\begin{equation}\label{eqn.SzRotation}
    R_{\bm{r}}^\alpha \equiv e^{-i \pi \alpha S^z_{\bm{r}}} = \cos{\left(\frac{\pi}{2} \alpha\right)} - 2 i S^z_{\bm{r}} \sin{\left(\frac{\pi}{2} \alpha\right)}~,
\end{equation}
which rotates the spin at the site $\bm{r}$ by $\pi \alpha$ counterclockwise around the $z$ axis in the spin space. 
In addition, we include perturbative couplings between wires for further optimization.

The lattice Hamiltonian we propose consists of two contributions, $H_\text{CSLL} = H_\diamondsuit(\alpha) + H^{\text{ZZ}}(J^z_\parallel,J^z_\perp)$, with
\begin{subequations}
\begin{align}
   H_\diamondsuit &= \frac{1}{2} \sum_{\bm{r}} \left[S^+_{\bm{r}} R^{\alpha \dagger}_{\bm{r} + \bm{b}_-} R^{\alpha}_{\bm{r} + \bm{b}_+}  S^-_{\bm{r}+\bm{b}_0}+ \text{H.c.}\right]~,\label{eqn.HD}\\
   H^{\text{ZZ}}&= J^z_\perp\sum_{\bm{r},i = \pm} S^z_{\bm{r}} S^z_{\bm{r} + \bm{b}_{i}}+ J^z_\parallel \sum_{\bm{r}} S^z_{\bm{r}} S^z_{\bm{r} + \bm{b}_{0}}~. \label{eqn.HZZ}
\end{align} 
\end{subequations}
Crucially, this Hamiltonian conserves $S^z$ on each wire separately. 
For noninteger $\alpha$, it contains three-spin terms such as $S_{\bm{r} + \bm{b}_+}^z \left(\vec{S}_{\bm{r}} \times  \vec{S}_{\bm{r}+\bm{b}_0} \right)^z$
that break time-reversal ${\cal T}$ and parity ${\cal P}$ symmetries, and are known to enhance quantum fluctuations and promote spin-liquid behavior~\cite{Wen89PRBa, Gong17PRB, Gorohovsky15prb, Cookmeyer21PRL}.

Below, in Sec.~\ref{sec.exact}, we show that $H_\text{CSLL}$ with $J^z_\perp=0$ maps exactly onto Eq.~\eqref{eqn.hxxz} via a non-local canonical transformation. On the long-wavelength degrees of freedom, this duality amounts to
\begin{align}
\varphi_y\rightarrow \varphi_y + \alpha \theta_{y+1} - \alpha \theta_{y-1},\quad \theta_y\rightarrow \theta_y, \label{eq.field.shift}
\end{align}
for $\alpha \in (-1,1)$. Consequently, for $J^z_\parallel \in (-1,1)$ and $|J^z_\perp| \ll 1$,  the long-wavelength description of $H_\text{CSLL}$ is given by
\begin{equation}\label{eqn.Fixedpoint}
\begin{split}
    \cH_{\text{CSLL}} &= \frac{uK}{2\pi} \sum_{y} \left(\grad \varphi_y + \alpha \grad \theta_{y+1} - \alpha \grad \theta_{y-1}\right)^2\\
    &+ \frac{u}{2\pi K}\sum_{y}\left[\left(\grad \theta_y \right)^2 + 2 \lambda \left(\grad \theta_y\right) \left(\grad \theta_{y+1}\right)\right]~.
\end{split}
\end{equation}
The parameters $u$ and $K$ are given by Eq.~\eqref{eqn.ba} as before and $\lambda = J^z_\perp \frac{K}{u} \frac{2}{\pi}$ according to Eq.~\eqref{eqn.Sz}.

Scaling dimensions at the fixed point $\cH_{\text{CSLL}}$ can be readily computed analytically. 
These quantities generally depend on the total number of chains $N$ and whether the transverse boundary condition is open or periodic. 
In the two-dimensional limit ($N \rightarrow \infty$), to the first order in $\lambda$, the scaling dimension of $\cO_{\text{chiral}}$ is 
\begin{align}
\label{eqn.deltachiral}
\Delta_{\text{chiral}} = &\frac{1}{2K} + 2K\left(1-\alpha+\frac{\alpha^2}{2}\right) \nonumber\\
&- \lambda \left[\frac{1}{4K} + K \left(1 - \frac{\alpha^2}{4}\right)\right]~.
\end{align}
Positive $\alpha$ and $\lambda$ reduce the scaling dimension, making $\cO_{\text{chiral}}$ more relevant at the low-energy scale. Even at $J^z_{\perp} = \lambda = 0$, interchain couplings parameterized by $\alpha$ reduce the lower bound on $\Delta_{\text{chiral}}$ from $2$ (where it is marginal) in the decoupled limit to $\sqrt{2}$ (where it is relevant). 

Notably, the scaling dimensions of $\cO_{\text{Ising}}$ and $\cO_{\text{VBS}}$ do not receive corrections from $\alpha$ or at the first order in $\lambda$ and follow Eq.~\eqref{eq.dimensions}. 
For the other two operators, the scaling dimensions up to the first order of $\lambda$ are
\begin{subequations}
\begin{align}
&\Delta_{\text{spiral}} = 1 + \frac{1}{2K} + K \alpha^2 - \frac{\lambda}{4} \left(\frac{1}{K} - K \alpha^2 \right)~,\\
&\Delta_{XY} = \frac{1}{2K} + 2 K \frac{3\alpha^2}{4}~.
\end{align}
\end{subequations}
We emphasize that all these scaling dimensions are perturbative in $\lambda$ but exact in $\alpha$.

The scaling dimension of $\cO_{\text{antichiral}}$ in Eq.~\eqref{eqn.spinexchange} follows from changing the sign of $\alpha$ in Eq.~\eqref{eqn.deltachiral}. In particular, it becomes less relevant for positive $\alpha$. However, we note that
\begin{equation}
\label{eqn.antichiral}
{\cal O}'_{\text{antichiral}} =\cos{\left(2\theta_{y-1} + 2\theta_{y+2} - \varphi_y + \varphi_{y+1}\right)}~
\end{equation}
realizes a CSL with the same chirality as ${\cal O}_{\text{antichiral}}$ and is favored by positive $\alpha$. Moreover, it is also generated at first order in $J_\perp$. At leading order in $\lambda$, its scaling dimension is related to that of ${\cal O}_{\text{chiral}}$ via
\begin{equation}
\Delta'_{\text{antichiral}} - \Delta_{\text{chiral}} = \lambda K \left(1-\alpha\right) ~.
\end{equation} 
At $\lambda = 0$, the two scaling dimensions coincide. Still, the two operators are unrelated by symmetries and thus generically occur with different bare coupling constants that will select one chirality over the other. In our work, we use positive $\lambda$ to reduce the value of $\Delta_\text{chiral}$, which additionally suppresses the opposite chirality.

\begin{figure}[tb]
    \includegraphics[width=\columnwidth]{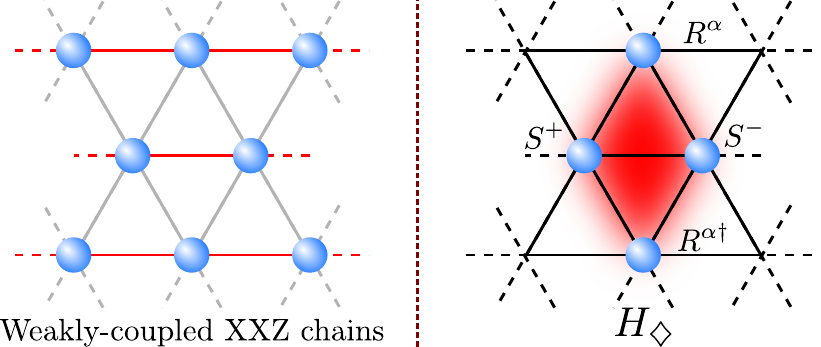}
    \caption{Comparison between weakly coupled $XXZ$ chains and Hamiltonian $H_\diamondsuit$ in Eq.~\eqref{eqn.HD}. For each red bond, the in-plane spin exchange interaction $S^+S^-$ is replaced by the product of the four operators at the corners of the red diamond in the right panel. The spin rotation operator $R^\alpha$ is a linear combination of $\mathbbm{1}$ and $S^z$ [cf.~Eq.~\eqref{eqn.SzRotation}], i.e., $H_\diamondsuit$ contains two-, three-, and four-spin interactions.}
    \label{fig.HD}
\end{figure}

\subsection{Integrability of the sliding Luttinger liquid}\label{sec.exact}

Two-leg ladder models similar to $H_\diamondsuit$ of Eq.~\eqref{eqn.HD} were introduced and studied in Refs.~\cite{Santra21PRB, Pozsgay22PRE}. Specifically, Ref.~\cite{Pozsgay22PRE} solved it using a nonlocal unitary transformation, akin to the Kennedy-Tasaki transformation~\cite{Kennedy92PRB, Oshikawa92}. 
Similarly, we introduce a set of nonlocal spin 1/2 operators $\tilde{S}$ with $\tilde S^z = S^z$ and
\begin{align} \label{Duality}
    \tilde{S}^-_{\bm{r}} &\equiv \left[\prod_{n = 1}^\infty R^\alpha_{\bm{r} - n\bm{b}_-} \right] S_{\bm{r}}^- \left[\prod_{n = 1}^\infty R^\alpha_{\bm{r} + n\bm{b}_-} \right]~,  \notag\\
    &\equiv \begin{bmatrix} 
    \cdots & R^\alpha & R^\alpha &              &                     & \\
           &          &          & S^-_{\bm{r}} &                     & \\
           &          &          &              & R^\alpha & R^\alpha & \cdots
    \end{bmatrix}~.
\end{align}
Here, each $R^\alpha$ is the rotation operator of Eq.~\eqref{eqn.SzRotation}. The second line illustrates the spatial structure of the string operators, which extend to $\pm \infty$ on the neighboring wires (all operators in the bracket are multiplied). The interaction in Eq.~\eqref{eqn.HD}, which includes terms on a `diamond' for four sites (Fig.~\ref{fig.HD}), maps onto an intrawire spin exchange in the dual variables, i.e.,
\begin{align} 
   \begin{bmatrix} 
                 & R^{\alpha}                      & \\
    S_{\bm{r}}^+ &                                                     & S_{\bm{r}+\bm{b}_0}^-\\
                 & R^{\alpha\dagger} &    
  \end{bmatrix} = \tilde{S}^+_{\bm{r}}  \tilde{S}^-_{\bm{r}+\bm{b}_0}~.
\end{align}
Consequently, Eq.~\eqref{eqn.HD} describes $XY$ chains of dual spins
\begin{align}
H_\diamondsuit[\Tilde{S}] = \frac{1}{2} \sum_{\bm{r}} \Tilde{S}^+_{\bm{r}} \Tilde{S}^-_{\bm{r}+\bm{b}_0} + \text{H.c.} ~\label{eqn.dualXY}
\end{align}
Since the spin $z$ component is unaltered in Eq.~\eqref{Duality}, the coupled array $H_{\text{CSLL}}$ of microscopic spins $S$ is thus equivalent to a decoupled, integrable array of dual spins $\Tilde{S}$ [Eq.~\eqref{eqn.hxxz}] when $|J^z_{\parallel}|<1$ and $J^z_{\perp} = 0$. Consequently, the long-wavelength behavior of this two-dimensional spin system is encoded in a sum of $\cH_{\text{LL}}$ in Eq.~\eqref{eqn.LLxxz} with non-local field variables $\tilde \theta_y,\tilde \varphi_{y}$, which are related to $\tilde S$ as in Eqs.~\eqref{eqn.Sz} and~\eqref{eqn.Sp}.

To obtain the effective field theory in Eq.~\eqref{eqn.Fixedpoint}, it remains to relate $\tilde \theta_y,\tilde \varphi_{y}$ to $ \theta_y, \varphi_{y}$. 
From here on, we take $\alpha \in [0,1)$ without loss of generality because $H_\diamondsuit$ is invariant under $\alpha \rightarrow \alpha+2$, and time-reversal relates positive and negative $\alpha$.
First, $S^z=\tilde S^z$ implies $\theta = \tilde \theta$. Second, bosonizing the string operator using the non-oscillating first term in Eq.~\eqref{eqn.Sz}, we get
\begin{align}\label{eqn.StringBoson}
    \prod_{m = 0}^\infty R^\alpha_{\bm{r} + m\bm{b}_0} 
    &= \exp\left[-i \pi \sum_{m = 0}^\infty \alpha S^z_{\bm{r} + m\bm{b}_0}\right] \\
    &\sim r_0 e^{-i \alpha \tilde{\theta}_y(x)}+ (-1)^x r_\pi e^{-i \left(\alpha - 2\right) \tilde{\theta}_y(x)},\nonumber
\end{align}
with nonuniversal constants $r_{0,\pi}$. For $\alpha <1$, which we assumed, we drop the second term as less relevant than the first. Inserting Eq.~\eqref{eqn.StringBoson} into Eq.~\eqref{Duality} and using Eq.~\eqref{eqn.Sp}, we conclude that
\begin{equation} \label{DualityField}
    \varphi_y \equiv \Tilde{\varphi}_y - \alpha \Tilde{\theta}_{y-1} + \alpha \Tilde{\theta}_{y+1}, \qquad \theta_y \equiv \Tilde{\theta}_y,
\end{equation}
as claimed in Eq.~\eqref{eq.field.shift}. Using this relation on an array of Luttinger liquids described by Eq.~\eqref{eqn.LLxxz} in $\tilde \theta,\tilde \varphi$, we arrive at Eq.~\eqref{eqn.Fixedpoint} with $\lambda=0$.

\begin{figure}[tb]
    \includegraphics[width=\columnwidth]{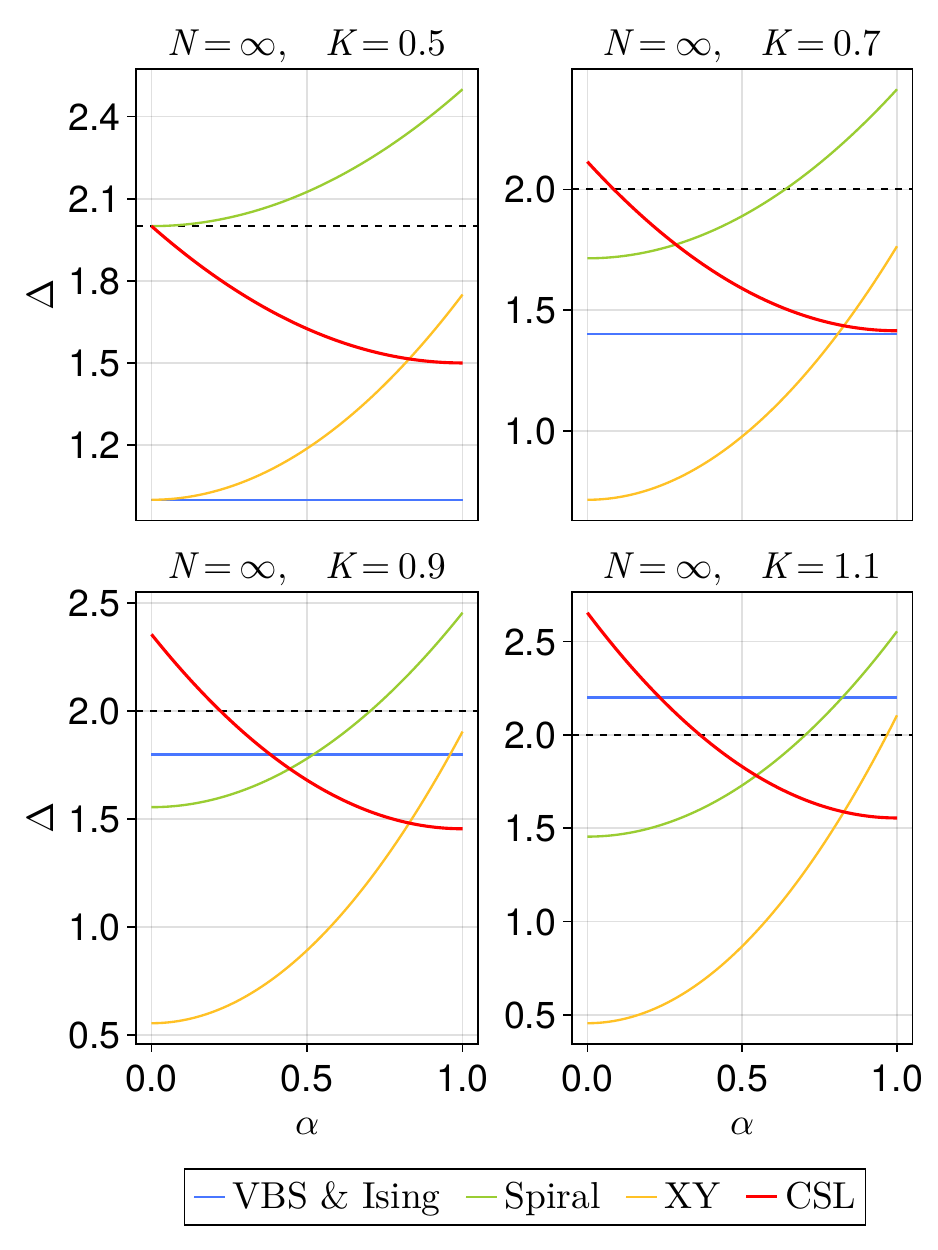}
    \caption{Scaling dimensions as functions of $\alpha$ in the two-dimensional limit ($N = \infty$) with $J^z_{\perp} = \lambda = 0$. In the two-parameter space of $K$ and $\alpha$, the operator $\cO_{\text{chiral}}$ can become a relevant perturbation ($\Delta_{\text{chiral}} < 2$). Moreover, the regime in which $\cO_{\text{chiral}}$ has the least scaling dimension among its competitors is finite. }
    \label{fig.ScalingsDuality}
\end{figure}

In Fig.~\ref{fig.ScalingsDuality}, we show the scaling dimensions of the CSL and other competitors as functions of $\alpha$ by fixing the parameter $K$. For additional details about the duality transformation and some numerical tests, see Appendix~\ref{app.duality}.

Physically, the duality transformation is closely related to flux attachment. 
Specifically, Eq.~\eqref{DualityField} is a truncated version of the transformation introduced in Refs.~\cite{Mross16PRLDirac, Mross17PRXDirac} to transform electrons or bosons into composite fermions. 
A crucial distinction is that the transformation used in the current work is only non-local in the $x$ direction and thus maps $H_\diamondsuit$ onto a local model. 
Still, the scaling dimensions in Eqs.~\eqref{eq.dimensions} and 
\eqref{eqn.deltachiral} show that, for sufficiently large $\alpha$, it favors ``composite fermion states'' such as the CSL---an integer quantum Hall state of composite fermions---over more conventional phases of the microscopic spins.

\subsection{Perturbative interwire couplings}

Perturbative interwire couplings can further optimize the SLL fixed point toward CSL instabilities.
One example is to include two-body spin-$z$ interactions in the vertical direction, i.e., the first term in Eq.~\eqref{eqn.HZZ}. In the perturbative regime where $J_\perp^z \ll 1$, using the bosonization dictionary Eq.~\eqref{eqn.Sz} immediately yields the last term in the low-energy effective Hamiltonian of Eq.~\eqref{eqn.Fixedpoint}. Beyond the perturbative regime, the correct fixed point includes additional symmetry-allowed terms, such as $\left(\partial_x \theta_y\right)\left(\partial_x \theta_{y+2}\right)$ \cite{Starykh07prl}.
Nevertheless, we expect Eq.~\eqref{eqn.Fixedpoint} to remain a reasonable approximation to the effective Hamiltonian until $J^z_\perp$ of order unity, which is confirmed by the numerical results shown in Appendix~\ref{app.psll}.

\begin{figure}[tb]
    \centering
    \includegraphics[width=\linewidth]{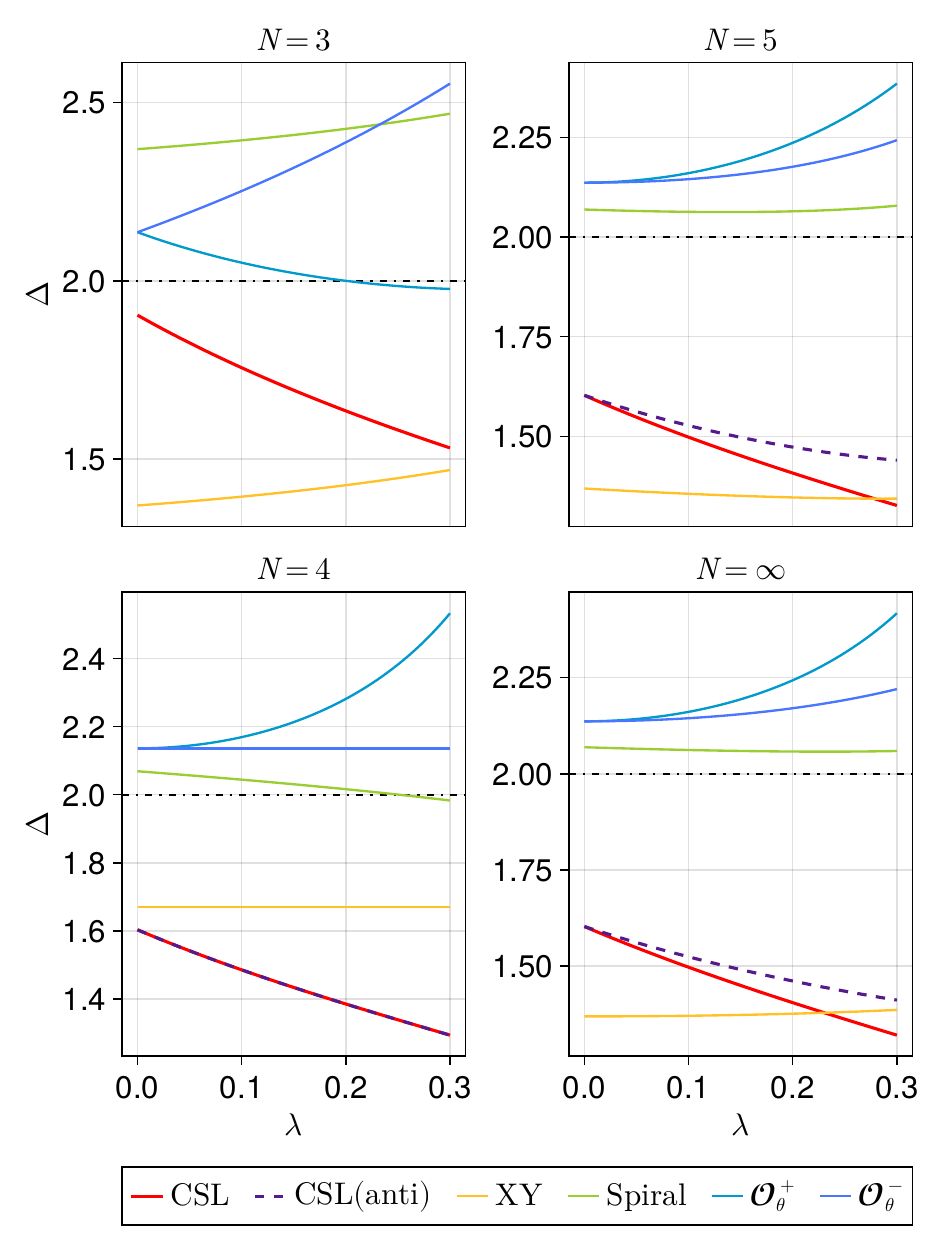}
    \caption{Scaling dimensions of various operators in systems with $N=3,4,5,\infty$ at $J^z_{\parallel} = -0.1$ and $\alpha = 0.75$. When the interwire density-density interaction is zero $(\lambda = 0)$, operators leading to CSLs do not have the smallest scaling dimension. At $J^z_\perp = 0.4$ ($\lambda \approx 0.3$), $\Delta_{\text{chiral}}$ is comparable to $\Delta_{XY}$ for small-width systems and is the smallest scaling dimension when $N = \infty$.}
    \label{ScalingD}
\end{figure}

The scaling dimensions at the fixed point defined by Eq.~\eqref{eqn.Fixedpoint} are shown in Fig.~\ref{ScalingD} for the choice of $J^z_\parallel = -0.1$ and $\alpha = 0.75$.
When the number of wires $N$ is finite, we implicitly assume a periodic boundary condition in the transverse direction.
We observe that the additional parameter $\lambda$ further reduces the scaling dimension of the ${\cal O}_\text{chiral}$ and separates it from its antichiral counterpart in the two-dimensional limit.
More details about calculating the scaling dimensions can be found in Appendix~\ref{app.scalings}.

\section{Precursors of the chiral spin liquid in quasi-one-dimension} \label{CSL1D}

Coupled-wire systems with a finite number of chains $N$, strictly speaking, are only quasi-one-dimensional. Therefore, no topological orders should occur \cite{Chen11PRB}. Nevertheless, the bosonized formulation provides a straightforward route for finding unique properties associated with the one-dimensional descendants of the CSL. In those phases, the arguments of the cosines in ${\cal O}_{\text{chiral}}$ are pinned to
\begin{equation}\label{eqn.CSLPinned}
  2 \theta_y + 2\theta_{y+1} + \varphi_y - \varphi_{y+1} = (2 n_y+1)\pi,
\end{equation}
with integer $n_y$, assuming a positive coupling for $\mathcal{O}_{\text{chiral}}$ in the field theory. (For the opposite sign in front of the cosine, the pinned values are even multiples of $\pi$). 
Since the system is periodic in the transverse direction, the equation implies
\begin{align}
\label{eqn.CSLPinned2}
  \Theta \equiv \sum_{y = 1}^{N} \theta_y = \frac{N}{4} \pi + \frac{\pi}{2} \sum_y n_y~.
\end{align}
Microscopic operators local in the $x$ direction (but not necessarily in $y$) must be invariant under the shift $\theta \rightarrow \theta +\pi$; see Eqs.~\eqref{eqn.Sz} and \eqref{eqn.Sp}. The simplest physical operator that acquires an expectation value when $\Theta$ is pinned is ${\cal O}_\Theta \equiv e^{2 i \Theta} $, which serves as an order parameter later on. Its two possible expectation values characterize distinct ground states for which the sum in Eq.~\eqref{eqn.CSLPinned2} is either odd or even.

\subsection{Spin pumping via flux insertion}
A hallmark of CSLs (FQHs) is that inserting a flux quantum is associated with accumulating a fractional spin (charge) excitation. We encode flux threading the cylinder through the boundary condition of the vertex field $e^{i \varphi_y}$ in the transverse direction. In particular, changing the flux by $\Phi$ amounts to
\begin{align}
  \varphi_N - \varphi_1 \rightarrow \varphi_N - \varphi_1+ \Phi ~.
\end{align}
Making this replacement in Eqs.~\eqref{eqn.CSLPinned} and \eqref{eqn.CSLPinned2}, we find that adiabatically inserting a full flux quantum, $\Phi=2\pi$, results in
\begin{align}
\delta\Theta_{2\pi}=\frac{\pi}{2}~.
\end{align}

The induced change of the spin in an interval $I_{R(L)}$ that encompasses the right (left) cylinder boundary but is smaller than the cylinder length is given by
\begin{align}\label{eqn.spinpump}
\delta S^z_{R(L)} &= \sum_{(x,y) \in I_{R(L)}} S^z_{x,y} \sim -\frac{1}{\pi} \int_{I_{R(L)}} \partial_x \delta\Theta_{2\pi}=\pm \frac{1}{2}~.
\end{align}
Here, we used $S^z$ of Eq.~\eqref{eqn.Sz} and that $\Theta$ outside the cylinder is pinned to a value that is insensitive to the flux insertion (Formally, one can model the trivial vacuum as an intrawire VBS state on each chain).

We emphasize that this argument does not make any assumption on $N$ and thus holds equally when $N \rightarrow \infty$. However, it does require a bulk energy density that is the same for $\Theta$ and $\Theta + \pi/2$. In two-dimensional limits, these two values encode different topological sectors. We now address the meaning of the two ground states in the quasi-one-dimensional regime.

\subsection{Degenerate ground states}
Our main findings are summarized in Table~\ref{XC2GroundStateSummary}. First, the two degenerate ground states may reveal themselves as symmetry-broken states in quasi-one-dimension, which can be inferred by the symmetry action on the vertex operator ${\cal O}_\Theta$. The symmetry properties of the vertex operator follow from the transformation of $\theta,\varphi$, summarized in Table~\ref{symmetries}. We find
\begin{align}
T_x {\cal O}_\Theta T_x^{-1}&= (-1)^N{\cal O}_\Theta \quad &&\text{(any $N$) }, \nonumber\\
\Pi {\cal O}_\Theta \Pi^{-1}&={\cal O}^\dagger_\Theta \quad &&\text{(any $N$) }, \label{eqn.syms}\\
 {\cal P T}{\cal O}_\Theta({\cal P T})^{-1}&=(-1)^{N/2}{\cal O}_\Theta \quad && \text{(even $N$ only)}, \nonumber
\end{align}
where the symmetry ${\cal P T}$ is only defined for an even number of wires with suitable boundary conditions (Fig.~\ref{Lattice}). 
In particular, the translation symmetry is broken for any odd $N$, while ${\cal P T}$ is broken for $N = 2,6,10,\ldots$.

When $N$ is an integer multiple of four, all symmetries are preserved in the ground state. We argue that two SPT states, possibly one trivial and one nontrivial, become degenerate in such arrays.
The symmetries in Eq.~\eqref{eqn.syms} imply that
\begin{align}
 {\cal H}_\Theta = g_\Theta\cos 2 \Theta
\end{align}
is an allowed perturbation in such arrays and selects a ground state with either $\Theta=0$ or $\Theta=\pi/2$, depending on the sign of $g_\Theta$. 
Microscopically, $g_\Theta$ arises from a string of spin operators wrapping around the cylinder and is thus exponentially suppressed in the circumference. 
Consequently, the energy splitting between the $\Theta=0$ and $\Theta=\pi/2$ states is also exponentially small. 

 The analysis done in Eq.~\eqref{eqn.spinpump} implies that domain walls between the two states trap a total spin $S^z_\text{DW} = \pm \frac{1}{2}$. Symmetry-protected boundary modes of this type are a signature of SPT phases, such as the Haldane phase \cite{Haldane83PRL, Affleck87VBS, denNijs89PRB, Kennedy92PRB, Kim99PRB}. We note that our lattice model $H_{\text{CSLL}} + H_{\perp}$ is invariant under rotation by $\pi$ along any two perpendicular axes in the spin space, and the same $\mathbb{Z}_2 \times \mathbb{Z}_2$ symmetry protects the Haldane phase in the spin-1 chain~\cite{Kennedy92PRB, Oshikawa92, Pollmann12PRB}.

\begin{table}[tb]
\begin{center}
\caption{\label{XC2GroundStateSummary} Ground states of the system on finite-width cylinders. Gray rows are obtained by deduction. The ground state properties exhibit a fourfold periodicity. Figure \ref{Lattice} shows the two different boundary conditions appearing in this table.}
\begin{tabular}{cccc}
\hline\hline
\textbf{Chains} & \textbf{Cylinder BC} & \textbf{Ground state} & \textbf{Simulated}\\
\hline
\textcolor{gray}{1} & \textcolor{gray}{NA} & \textcolor{gray}{$T_x$-broken} & \\ 
\textcolor{gray}{2} & \textcolor{gray}{XC-2(0)} & \textcolor{gray}{$\mathcal{P} \mathcal{T}$-broken} & \\
3          & XC-3 & $T_x$-broken & \cmark\\ 
4          & XC-4(0) &SPT (Haldane)  & \cmark\\ 
5          & XC-5 & $T_x$-broken & \cmark\\ 
6          & XC-6(0) & $\mathcal{P} \mathcal{T}$-broken & \cmark\\ 
7          & XC-7 & $T_x$-broken &    \\ 
8          & XC-8(0) &SPT&    \\ 
\vdots       & \vdots & \vdots & \\
$\infty$       & & CSL & (\cmark) \\
\hline\hline
\end{tabular}
\end{center}
\end{table}
\subsection{String order parameters}

Nonlocal observables involving string operators discussed in Eq.~\eqref{eqn.StringBoson}, i.e.,
\begin{equation}\label{eqn.singlestring}
  s^\beta_{y}(x) \equiv \prod_{m = 0}^x R^\beta_{m\bm{b}_0}\sim e^{-i \beta \theta_y(0) + i \beta \theta_y(x)}
\end{equation}
can provide additional evidence of a CSL ground state. In particular, we find that the product of the operators over all wires with cylindrical boundary conditions satisfies
\begin{equation}
\begin{split}
 \expval{\prod_y s^\beta_{y}(x)} \sim \expval{e^{-i \beta \Theta(0)+i \beta \Theta(x)}} = e^{-\frac{\beta^2}{2} \expval{ \left[\Theta(0)-\Theta(x)\right]^2}}~,
\end{split}
\end{equation}
which has a finite $x\rightarrow \infty$ limit for the CSL since $\Theta$ is pinned. However, since the system is fully gapped, the expectation value of each individual $s_y^\beta$ (or any of their products) decays exponentially unless $\theta_y$ is pinned, i.e., unless the $y$th chain is in a VBS or Ising state. The analysis shows that expectation values of different string operators can provide a clear signature of quasi-one-dimensional CSL precursors.

For open boundary conditions in the $y$ direction, Eq.~\eqref{eqn.spinexchange} contains one fewer cosine than for periodic boundaries. Consequently, Eq.~\eqref{eqn.CSLPinned} holds for $N-1$ terms only. Summing these terms, we find
\begin{equation}
 4 \Theta =\sum_{y=1}^{N-1} (2 n_y+1)\pi +( 2 \theta_N + \varphi_N+ 2\theta_{1} - \varphi_{1}).
\end{equation}
The additional terms on the right-hand side describe gapless edge excitations at the $y=1$ and $y=N$ boundaries. 
These edge modes result in a universal power law decay of the spin-spin correlations along the boundary, given by $\langle S^+_{0,y} S^-_{x,y}\rangle \sim x^{-2}$ for $y=1$ or $y=N$. 
Similarly, we find that the string operator for a system with open boundaries exhibits a slow power-law decay 
\begin{equation}
\begin{split}
 \expval{\prod_y s^\beta_{y}(x)}_\text{OBC}\sim \frac{1}{x^{\beta^2/4}}~.
\end{split}
\end{equation}

\section{Tensor network simulations} \label{Numerics}

The lattice Hamiltonian for realizing the CSL is a sum of $H_{\perp}$ of Eq.~\eqref{eqn.hperp} and $H_{\text{CSLL}}$ of Eqs.~\eqref{eqn.HD} and \eqref{eqn.HZZ}. For our simulations, we choose the parameters
\begin{align}\label{eqn.Choice}
    \alpha = 0.75~, \quad J^z_{\parallel} = -0.1~, \quad 
    J^z_{\perp} = 0.4~
\end{align}
in $H_{\text{CSLL}}$.
This choice corresponds to $\lambda \approx 0.3$, for which $\cO_{\text{chiral}}$ is the most relevant perturbation when $N \geq 4$; see Table~\ref{ScalingdimensionTable} and Appendix~\ref{app.scalings}.
The final coupling we took is $J_\perp=0.28$, a compromise between analytical control (favoring small $J_\perp$) and numerical practicality (dictating larger $J_\perp$). At $N=3$, $\cO_{\text{chiral}}$ stands in close competition with a magnetically ordered state but is favored by a parametrically larger bare coupling $\propto J_\perp$ versus $J_\perp^2$ for its competitor.

We numerically tested for the CSL signatures discussed in Sec.~\ref{CSL1D}. For this purpose, we wrapped the lattice Hamiltonian around cylinders under different boundary conditions depicted in Fig.~\ref{Lattice} and simulate this quasi-one-dimensional system with circumference $N=3,4,5,6$ using finite-size DMRG and VUMPS achieved by the ITensor library \cite{ITensor, ITensor-r0.3}. All simulations take advantage of the total $S^z$ conservation and focus on the sectors with total $S^z=0$ or $\frac{1}{2}$.

\setlength{\extrarowheight}{4pt}
\begin{center}
\begin{table}[tb]
\caption{\label{ScalingdimensionTable} Scaling dimensions to the second digit, evaluated at $J^z_{\parallel} = -0.1$, $\alpha = 0.75$, and $\lambda = 0.3$ at the SLL of Eq.~\eqref{eqn.Fixedpoint}.
}
\begin{tabularx}{\linewidth}{ccccccc}
\hline\hline
                     & $\Delta_{\text{chiral}}$  & $\Delta'_{\text{antichiral}}$  & $\Delta_{XY}$  & $\Delta_{\text{spiral}}$ & $\Delta_{\theta}^+$
                     & $\Delta_{\theta}^-$\\
\hline
$N=3$                & $1.53$ &        & $1.47$ & $2.47$ & $1.98$ & $2.55$\\
$N=4$                & $1.29$ & $1.29$ & $1.67$ & $1.98$ & $2.53$ & $2.14$\\
$N=5$                & $1.33$ & $1.44$ & $1.34$ & $2.08$ & $2.38$ & $2.24$\\
$\vdots$             &$\vdots$&$\vdots$&$\vdots$&$\vdots$&$\vdots$& $\vdots$\\
$N\rightarrow\infty$ & $1.32$ & $1.41$ & $1.39$ & $2.06$ & $2.42$ & $2.22$\\
\hline\hline
\end{tabularx}
\end{table}
\end{center}

\subsection{Spin pumping}

\begin{figure}[tb]
    \centering
    \includegraphics[width=\linewidth]{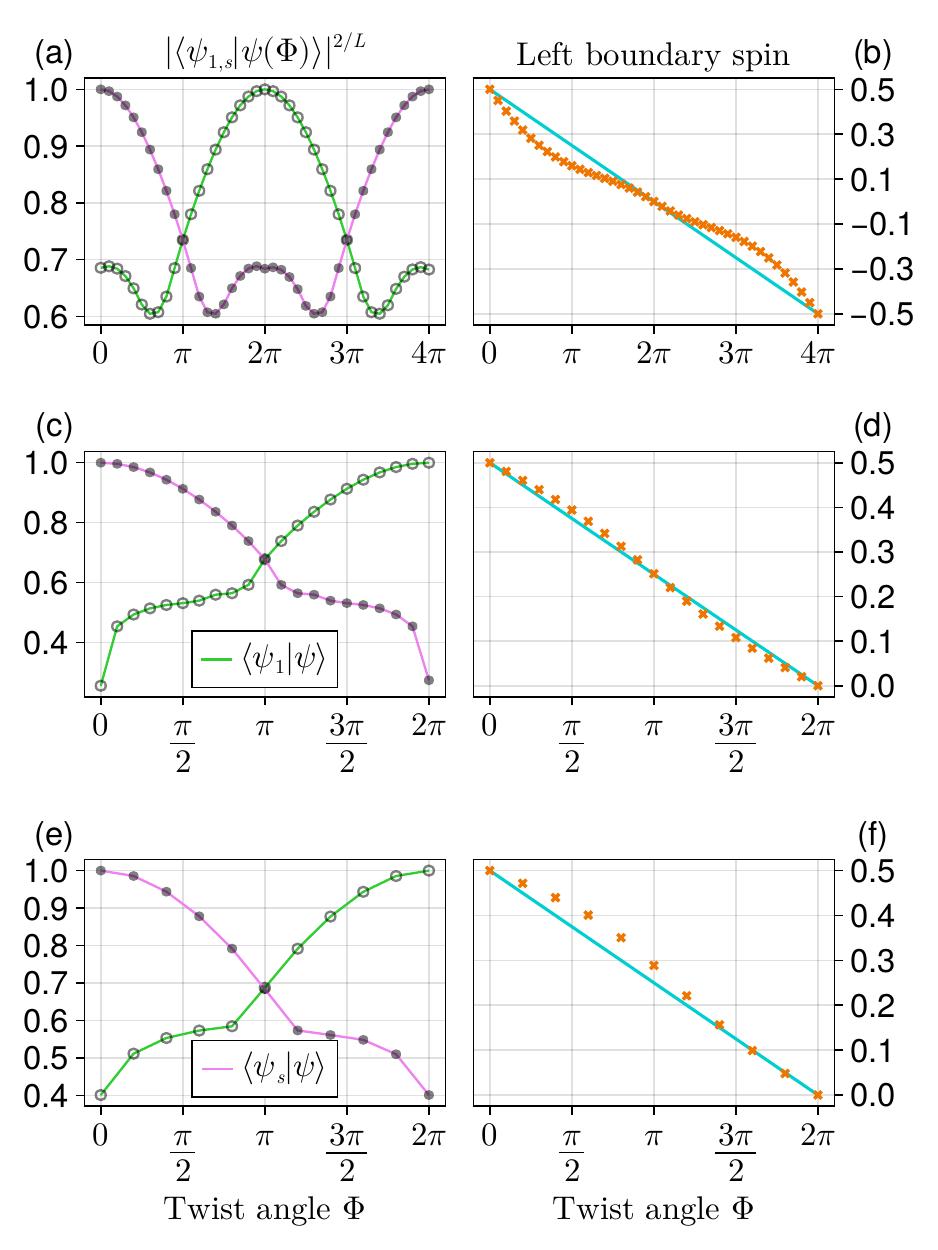}
    \caption{Flux insertion simulations for various systems with different numbers of chains. The top row shows data for the XC-3 cylinder obtained from infinite MPSs ($L = \infty$) at $\chi = 1500$. Panel (a) shows the fidelity per unit cell (black markers) throughout the insertion between the infinite MPS and the two degenerate ground states $\psi_1$ and $\psi_s$. Panel (b) shows how much spin stays at the left (infinite) end of the cylinder when flux insertion is performed in orange crosses. The blue line is the theoretical curve, assuming the system is a CSL.  The central row shows data obtained from MPSs at $\chi = 2000$ for the open XC-5 cylinder with $31$ unit cells. Panel (c) monitors $|\braket{\psi_{1,s}}{\psi(\Phi)}|^{2/L}$, similar to the fidelity per unit cell for an infinite MPS. Panel (d) shows the boundary spin. The bottom row shows data obtained from MPSs at $\chi = 2000$ for the open XC-6(0) cylinder with $32$ unit cells.}
    \label{fig.SpinPumping}
\end{figure}

We start to present numerical results from the spin-pumping property. For the descendant of the CSL in one dimension, a $2\pi$-flux insertion shifts the ground state to the other (degenerate) state with a different topological flux when the number of chain $N$ is sufficiently large, as discussed in Sec.~\ref{CSL1D}.
Simultaneously, the system transfers $S^z = \pm 1/2$ from the cylinder's left end to the right end. 
The boundary spin value can be easily obtained from matrix product states (MPSs) describing finite systems. 
For an infinite MPS, we probe the boundary spin by looking at the spectrum of reduced density matrices of half-infinite cylinders~\cite{Zaletel14JSM}. 
The reduced density matrix $\rho_L$ of the left half-infinite cylinder is obtained by partially tracing out the sites with $x>0$ from the density matrix $\dyad{\psi}{\psi}$, where $\ket{\psi}$ is the infinite MPS.
Each eigenvector of $\rho_L$ corresponding to the eigenvalue $p_j$ can also be a simultaneous eigenvector of $\sum_{\bm{r} \in L} S^z_{\bm{r}}$ with eigenvalue $S^z_j$.
Thus, the left boundary spin is
\begin{equation}
    S^z_L = \sum_j p_j S^z_j~.
\end{equation}

The correct spin-pumping features emerge in the three-chain systems. First, we find two degenerate ground states for the XC-3 cylinder with the same energy density $(e = E/L \approx -0.9985)$ within numerical accuracy. The two states differ by whether there is a boundary spin $1/2$ ($\psi_s$) or not ($\psi_1$) for the left half-infinite system. 
To insert flux, we use the twisted boundary condition in the transverse direction and follow the adiabatic procedure of Refs.~\cite{He14PRB, He14prl, He15PRL, He17PRX, Szasz20PRX}. 
Specifically, we increase the flux $\Phi$ through the cylinder in small increments of $\pi/10$, taking the state at the previous $\Phi$ as the initial state for the current iteration. 
Starting with the infinite MPS $\psi_s$ with boundary spin $S^z_{L}=1/2$, we show the boundary spin of $\psi(\Phi)$ in Fig.~\ref{fig.SpinPumping}(b). 
As anticipated, changing the flux by $2\pi$ eliminates the boundary spin. 
Increasing the flux by another $2\pi$ changes the boundary spin to $S^z_{L}=-1/2$.

Additionally, in Fig.~\ref{fig.SpinPumping}(a), we monitor the fidelity per unit cell as a measure of how close the infinite MPS $\psi(\Phi)$ is to $\psi_{1,s}$ in the bulk of the system ~\cite{Vanderstraeten19SciPost}. 
A unit fidelity does \textit{not} imply that two infinite MPSs are identical. In particular, the fidelity is insensitive to boundary effects, such as the residual spin, encoded in the MPSs.
For further information about the fidelity, see Appendix~\ref{app.tensors}. As the twist angle increases, the fidelity between $\psi(\Phi)$ and $\psi_s$ decreases, i.e., the two states are orthogonal in the thermodynamic limit. 
After a $2\pi$ flux insertion, the fidelity between $\psi(2\pi)$ and the second state, $\psi_1$, reaches unity. 
A second flux insertion recovers the original fidelity with $\psi_s$.

In the case of $N=5$ and $N=6$, VUMPS did not converge for the chosen maximum bond dimension, which we attribute a large entanglement between neighboring unit cells representing adjacent rings on the cylinder. In this case, we observed a better-behaved convergence using finite-size cylinders, which we modeled using a finite MPS ansatz combined with DMRG.
At $N=5$, we choose an open XC-5 cylinder with $L=31$ unit cells and focus on the $\sum_i S^z_i = 1/2$ sector. Choosing an odd $L$ allows the system to have two degenerate ground states while the ground state is nondegenerate for an even and finite $L$. For six-chain systems, we use the XC-6(0) boundary condition and pick $L=32$. Similar to the three-chain case, we first prepare the state to have a spin of $1/2$ at the left boundary, which corresponds to $\Phi = 0$ of Figs.~\ref{fig.SpinPumping}(d) and \ref{fig.SpinPumping}(f). A $2\pi$-flux insertion eliminates the boundary spin, and the data across the flux insertion sit closely to the theoretical curve in blue.

In contrast, we do not find the same striking features for $N=4$ cylinders. According to Table \ref{XC2GroundStateSummary}, the ground state is nondegenerate for four-chain systems. 
In practice, we find that the other supposedly degenerate state is embedded within the excited states, which impedes the flux insertion from swapping the two topological sectors.
The same spin-pumping property should emerge once we make $N$ large enough, yet the next relevant value, $N=8$, is still out of our simulation capacity.
Consequently, we utilize other approaches to confirm the occurrence of the CSL in four-chain systems, as elaborated below.

\subsection{Entanglement signatures}

The entanglement spectrum of a CSL exhibits a universal pattern that can serve as a fingerprint \cite{Li08PRL}. Specifically, we consider the XC-N transverse boundary condition to maximize the allowed momentum values and introduce a real-space partition that separates the cylinder into two half-infinite subsystems. Due to the periodic boundary condition, the spectrum of the reduced density matrix for the left half-infinite system can be labeled by the transverse momentum $k_y \in \{0,2\pi/N,4\pi/N, \cdots, 2\pi\left(1-1/N\right)\}$ in addition to the U(1) quantum number $S^z$. 
For the Kalmeyer-Laughlin CSL, this spectrum is expected to show several distinct features when $N$ is sufficiently large \cite{Moore97PRB, ArildsenPRB22}:
\begin{enumerate}
    \item {\bf Lowest eigenvalue}---The lowest eigenvalue is nondegenerate if the state is in the identity sector, while it is two-fold degenerate for states in the semion sector. In particular, the lowest eigenvalue occurs in two consecutive $U(1)$ quantum number sectors when it is doubly degenerate.
    \item {\bf $\bm{U(1)}$ symmetry}---The spectrum is symmetric around $S^z = j_0$ for some $j_0$. Specifically, for every eigenvalue $E_j(k_y)$, $E_{2j_0 - j}(k_y)$ is also an eigenvalue of the spectrum. Consequently, the lowest eigenvalue occurs in the $S^z = j_0$ sector (identity) or the $S^z = j_0 \pm 1/2$ sectors (semion).
    \item {\bf Degeneracy pattern}---The low-lying eigenvalue levels in each $S^z$ sector correspond to one branch of chiral bosonic excitations governed by the $U(1)$ Kac-Moody algebra \cite{Wen92}. In particular, their degeneracies $\{1,1,2,3,5,7,11,\cdots\}$ are given by partitions of integers. Furthermore, the eigenvalue increases uniformly between consecutive momenta, quantized in units of $2\pi/N$ \footnote{Without loss of generality, we pick $\delta k_y > 0$ by fixing a chirality}. 
    \item {\bf Energy and momentum offsets}---Up to a global shift and rescaling, the spacing between consecutive levels in each $U(1)$ quantum number sector is $1$, and the lowest eigenvalue in the entire spectrum is $0$ ($1/4$) for states in the identity (semion) sector. 
    The momentum of the lowest level in the $S^z = j$ sector is
    \begin{equation}\label{eqn.IdentityMomentum}
        k_{1}^{(j)} = k_{1}^{(j_0)} + \frac{2\pi}{N}\left(j - j_0\right)^2
    \end{equation}
    for states in the identity sector and is 
    \begin{equation}\label{eqn.SemionMomentum}
        k_{1}^{(j)} = k_{1}^{(j_0 \pm 1/2)} + \frac{2\pi}{N} \left[\left(j - j_0\right)^2 - \frac{1}{4}\right]
    \end{equation}
    for states in the semion sector.
\end{enumerate}

\begin{figure}[hbt]
\includegraphics[width=\columnwidth]{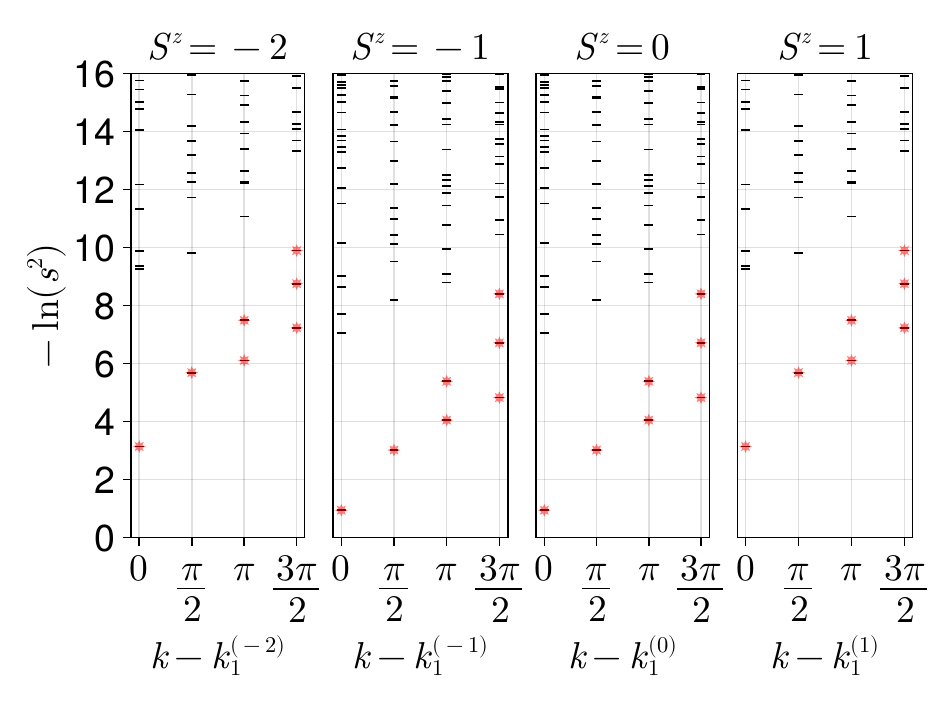}
\caption{
Momentum-resolved entanglement spectrum evaluated at the ground state of the XC-4 cylinder with bond dimension $\chi = 1500$. The ground state has a semionic charge since the lowest eigenvalues occur twice in the $S^z = -1,0$ sectors. The horizontal axis for each panel represents $k - k_1^{(j)}$, where $k_1^{(j)}$ is defined in Eq.~\eqref{eqn.SemionMomentum}.
The low-lying part of the levels in each sector shows a clear degeneracy pattern of $\{1,1,2,3\}$.} 
\label{FourChainES}
\end{figure}

Here, we show results from four- and five-chain cylinders. The VUMPS algorithm yields one state for four-chain systems under the XC-4 transverse boundary condition. As shown in Fig.~\ref{FourChainES}, the lowest eigenvalue is doubly degenerate and occurs in $S^z = -1,0$ sectors, which means the state carries topological flux. It is further confirmed by the symmetric pattern shown in the spectrum. Moreover, the levels in each quantum number sector exhibit a clear chirality and follow the expected degeneracy pattern for a Laughlin state, i.e., $1,1,2,3,\cdots$ \cite{Wen92, Moore97PRB, Li08PRL}.

\begin{figure}[tb]
    \centering
    \includegraphics[width=\linewidth]{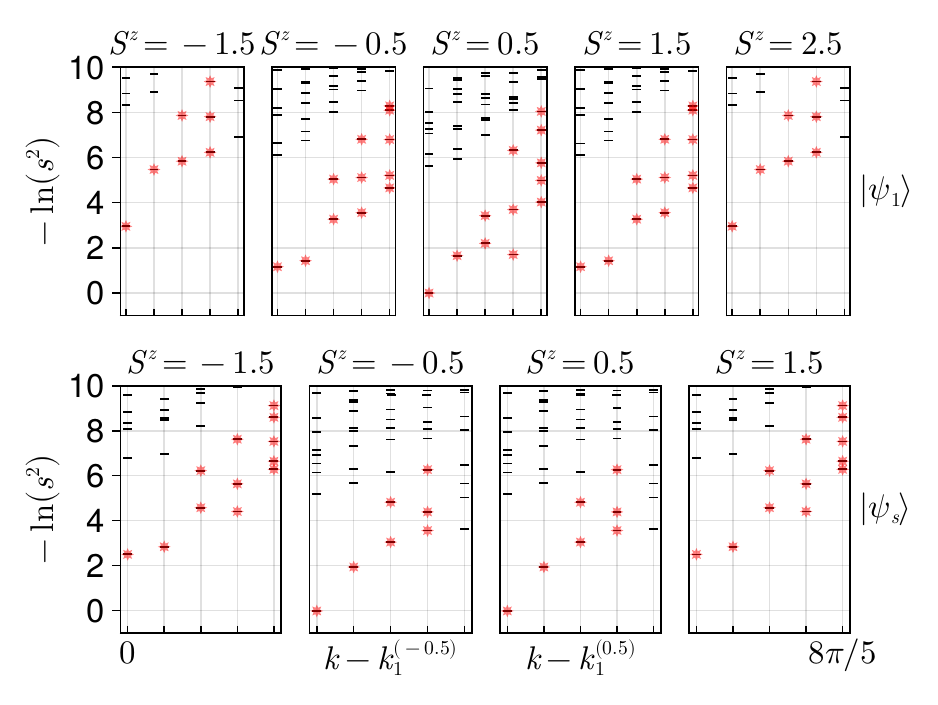}
    \caption{Momentum-resolved entanglement spectra of MPSs at $\chi = 2000$ for a five-chain finite cylinder with $31$ unit cells. We first shift the spectra by a constant such that the lowest eigenvalue is $0$. The five upper figures describe state $\psi_1^{\text{MPS}}$ in the identity sector, while the state $\psi_s^{\text{MPS}}$ yields the four lower figures. The horizontal axis for each panel represents $k - k_1^{(j)}$ where $k_1^{(j)}$ is defined in Eqs.~\eqref{eqn.IdentityMomentum} and \eqref{eqn.SemionMomentum}. Ticks on the axis for each panel are $0, 2\pi/5, \cdots, 8\pi/5$ from left to right. 
    Both states reveal the correct degeneracy $(1,1,2,3,5)$ of the low-lying eigenvalues.}
    \label{FiveChainES}
\end{figure}

We have observed similar properties of the entanglement spectra when $N=5$. As mentioned in the previous subsection about spin pumping simulations, we utilize DMRG to simulate open cylinders with $31$ unit cells. Throughout the simulation, we cut the entire system so that the left half always has $16$ unit cells. Consequently, the MPS with a spin of $1/2$ sitting at the left end should yield another spinon at the virtual boundary and, thus, is regarded as a state $\psi_s$ in the semion sector. Similarly, the other degenerate MPS $\psi_1$ with a spin of $1/2$ sitting at the right end corresponds to the identity sector.
Figure \ref{FiveChainES} shows the momentum-resolved spectra of the two degenerate MPSs. The spectra are consistent with our assignment for the topological sector. Within each $U(1)$ quantum number sector, we have marked a few low-lying eigenvalues that obey the correct conformal field theory counting $\{1,1,2,3,5\}$.

In six-chain cylinders, we find entanglement spectra whose counting is consistent with expectations but which are more strongly affected by the finite bond dimension or the system size. They are shown in Appendix~\ref{app.numerics} along with spectra for the three-chain system.

\subsection{Quasi-one-dimensional signatures}

To obtain additional evidence of CSL behavior, we use the other signatures derived in Sec.~\ref{CSL1D}. Specifically, we argued that the topologically degenerate states reveal themselves by spontaneously breaking certain symmetries or by realizing SPT phases in quasi-one-dimension. In addition, the CSL precursors are characterized by specific non-local string order parameters.

\begin{figure}[htb]
    \centering
    \includegraphics[width=\linewidth]{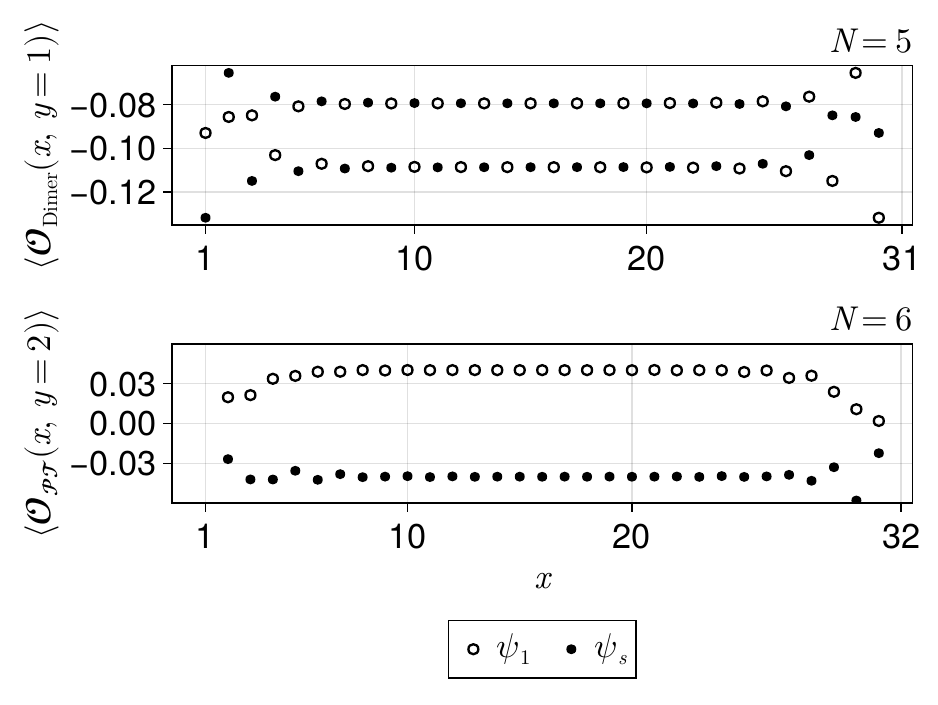}
    \caption{
    Spontaneous symmetry breaking in $N=5$ and $N=6$ systems. The hollow (filled) points correspond to the state in the identity (semion) sector. All MPSs have a maximum bond dimension $\chi = 2000$. Upper panel: the expectation values of $\mathcal{O}_{\text{Dimer}}$ on the XC-5 cylinder with $31$ unit cells. The values oscillate in the bulk of the cylinder, signaling a spontaneous symmetry breaking in translation. Lower panel: the expectation values of the order parameter $\mathcal{O}_{\mathcal{P}\mathcal{T}}$ on the XC-6(0) cylinder with $32$ unit cells. $\expval{\mathcal{O}_{\mathcal{P}\mathcal{T}}} \neq 0$ in the bulk means both $\psi_1$ and $\psi_s$ break the $\mathcal{P}\mathcal{T}$ symmetry.}
    \label{fig.TSSB}
\end{figure}

According to Table~\ref{XC2GroundStateSummary}, the system breaks $x$-translation symmetry when the total number of chains $N$ is odd and $\mathcal{P}\mathcal{T}$ symmetry when $N/2$ is odd. We tested for spontaneous symmetry breaking of translation symmetry by looking at the dimer operator
\begin{equation}\label{eqn.TOrderParam}
    \expval{\cO_{\text{Dimer}}(\bm{r}+\bm{b}_0/2)} \equiv \expval{\vec{S}_{\bm{r}} \cdot \vec{S}_{\bm{r} + \bm{b}_0}}~.
\end{equation}
For a translationally invariant state along $\bm{b}_0$, this quantity is uniform; an oscillating $\expval{\cO_{\text{Dimer}}}$ thus implies a spontaneous symmetry breaking. The two expectation values in Fig.~\ref{fig.TSSB} relate to each other by the $x$-translation in the bulk, which is consistent with the spontaneous symmetry breaking.

The spontaneous $\mathcal{P}\mathcal{T}$ symmetry breaking for XC-N(0) cylinders when $N/2$ is odd manifests in the dimer operator in the transverse direction. Specifically, we define
\begin{equation}\label{eqn.PTOrder}
    \begin{split}
        \mathcal{O}_{\mathcal{P}\mathcal{T}}(\bm{r}) &\equiv \vec{S}_{\bm{r}} \cdot \vec{S}_{\bm{r} - \bm{b}_+} - \vec{S}_{\bm{r}} \cdot \vec{S}_{\bm{r} + \bm{b}_-}~.
    \end{split}
\end{equation}
This operator measures the difference of the two dimers and is odd under $\mathcal{P}\mathcal{T}$. Therefore, it must have zero expectation value for a $\mathcal{P}\mathcal{T}$ symmetric state.
Nevertheless, the expectation value is non-zero in the bulk as shown in Fig.~\ref{fig.TSSB}, indicating a spontaneous $\mathcal{P} \mathcal{T}$ symmetry breaking for both $\psi_1$ and $\psi_s$. We further note that the data also support $\mathcal{P} \mathcal{T} (\psi_1) = \psi_s$, another signature of spontaneous symmetry breaking.

\begin{figure}[tb]
\includegraphics[width=\columnwidth]{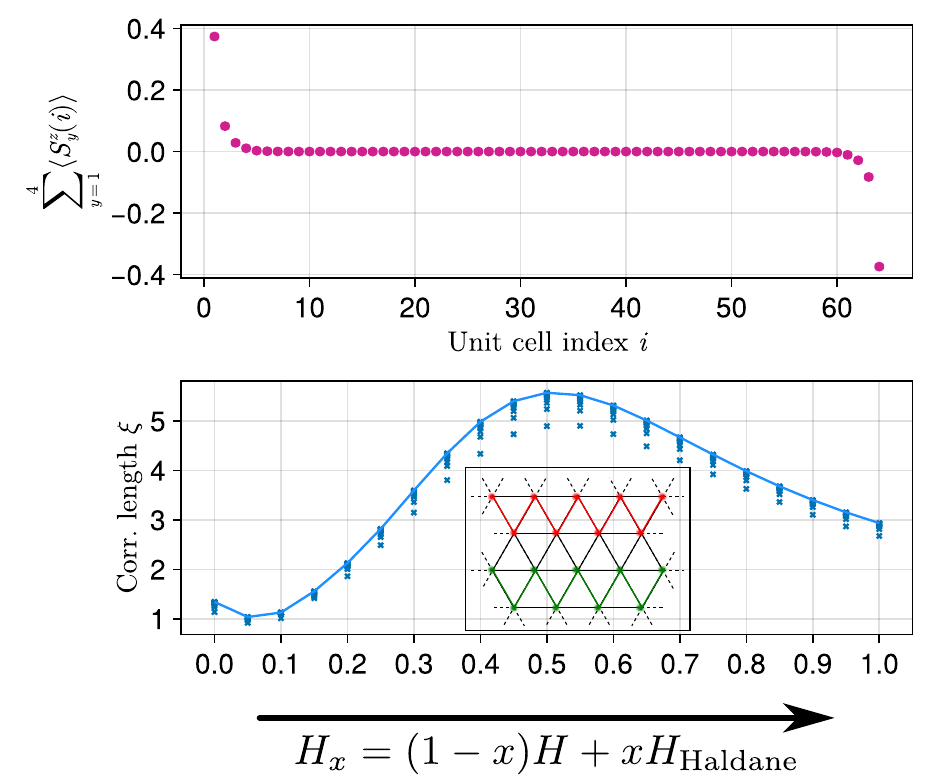}
\caption{Upper figure: Expectation values of $\sum_{y=1}^4 S^z_{y}(i)$ versus the unit cell index $i$. The system contains $64 \times 4$ spin 1/2 sites under the XC-4(0) boundary condition. The MPS has a maximal bond dimension of $1000$. The total spin localized at the left (right) boundaries is $0.5$ ($-0.5$). Lower figure: Correlation lengths versus the interpolation parameter $x$. The Hamiltonian is our lattice model with $N = 4$ at $x = 0$. At $x = 1$, it is the model $H_\text{Haldane}$ defined in Eq.~\eqref{DiagonalLadder} with SO(3) spin rotation symmetry. Each data point is obtained at a given $x$ and a bond dimension within $\{100, 200, \cdots, 1000\}$. The solid curves consist of points with bond dimensions $1000$. The inset figure sketches how $H_\text{Haldane}$ effectively describes a two-leg spin ladder. Sites with the same color belong to one leg.} 
\label{Haldane}
\end{figure}

In four-chain ($N=4$) systems, there is no spontaneous symmetry breaking according to Table~\ref{XC2GroundStateSummary}. 
Indeed, we find a unique ground state $\psi_s$ being in the semion sector for the XC-4(0) cylinder using the VUMPS algorithm. 
When we consider a finite open cylinder, the corresponding ground state from DMRG shows spin 1/2 degrees of freedom at the boundaries (Fig.~\ref{Haldane}). 
Consequently, we suspect the symmetric ground state is adiabatically connected to the Haldane phase \cite{Haldane83PRL, Affleck87VBS}. 
To confirm our conjecture, we design an interpolation between our Hamiltonian on the XC-4(0) cylinder and
\begin{equation}\label{DiagonalLadder}
    \begin{split}
    H_\text{Haldane} &= \sum_j \left(\vec{S}_{j-\frac{1}{2},1} + \vec{S}_{j-\frac{1}{2},3}\right) \cdot \left(\vec{S}_{j,2} + \vec{S}_{j,4}\right)\\
    &+ \sum_j \left(\vec{S}_{j,2} + \vec{S}_{j,4}\right) \cdot \left(\vec{S}_{j+\frac{1}{2},1} + \vec{S}_{j+\frac{1}{2},3}\right)~,
    \end{split}
\end{equation}
which can be viewed as a two-leg spin ladder (see the inset in Fig.~\ref{Haldane}). The model has been shown to lie in the Haldane phase in Refs. ~\cite{Strong94PRB, Kim99PRB}. The linear interpolation preserves all the symmetries in our model because they are also symmetries in $H_{\text{Haldane}}$. The correlation lengths of the infinite MPSs during the interpolation are shown in Fig.~\ref{Haldane}. The solid dots are data obtained from infinite MPSs at different bond dimensions, and the lines connect the data from the largest bond dimension. Since the correlation length does not diverge, we conclude that the two phases are adiabatically connected.

\begin{figure}[tb]
    \includegraphics[width=\columnwidth]{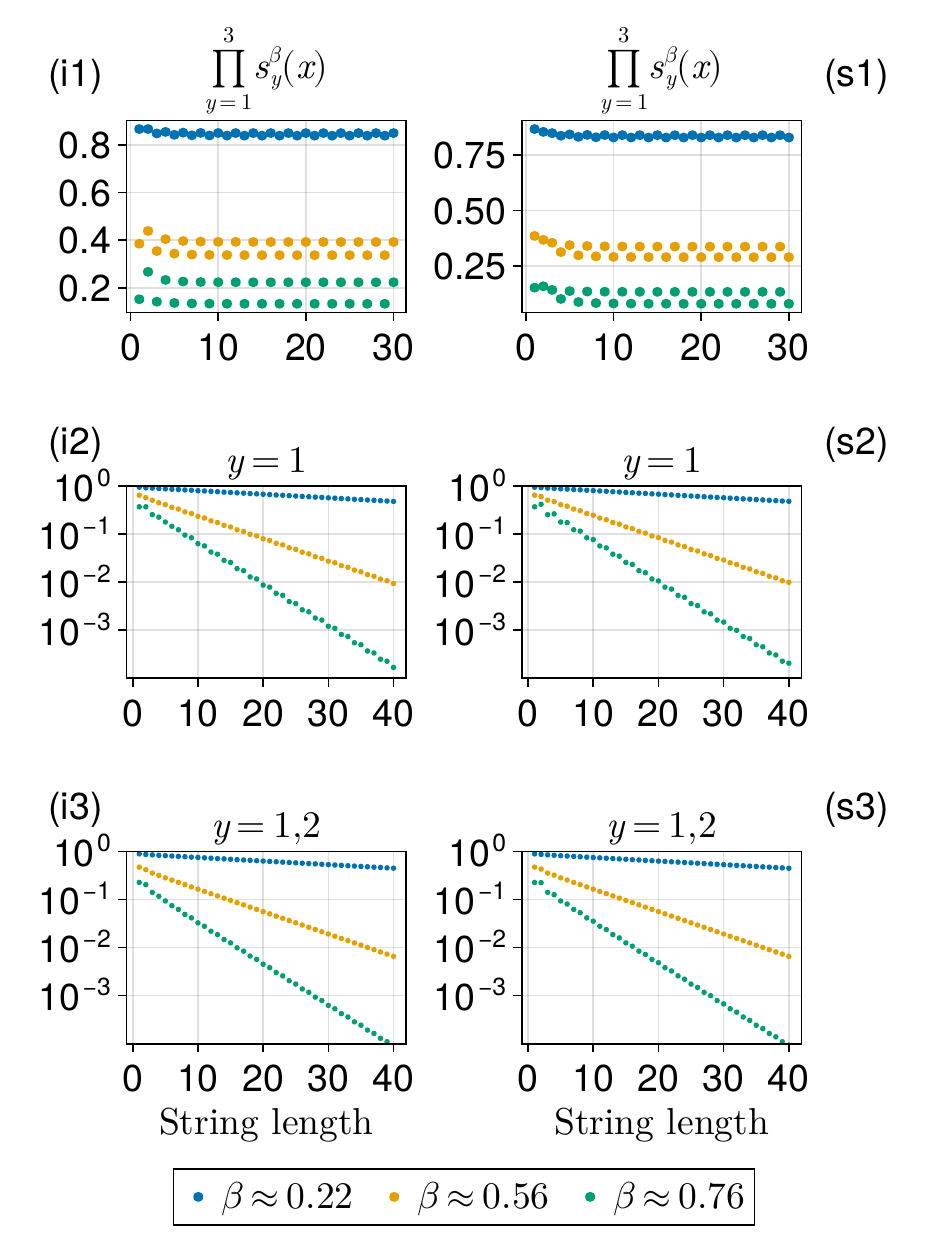}
    \caption{Expectation values for different string operators evaluated by the infinite MPS in the identity sector (i1--i3) or the semion sector (s1--s3) at bond dimension $\chi = 2000$ on the XC-3 cylinder. (i1, s1) When the product goes through all the chains, the corresponding operator develops a nonzero expectation value for $x\gtrsim 10$. (i2, i3, s2, s3) If the string operator does not involve all three chains, the expectation value decays exponentially as the length increases.} 
    \label{ThreeChainString}
\end{figure}

\begin{figure}[tb]
    \includegraphics[width=\columnwidth]{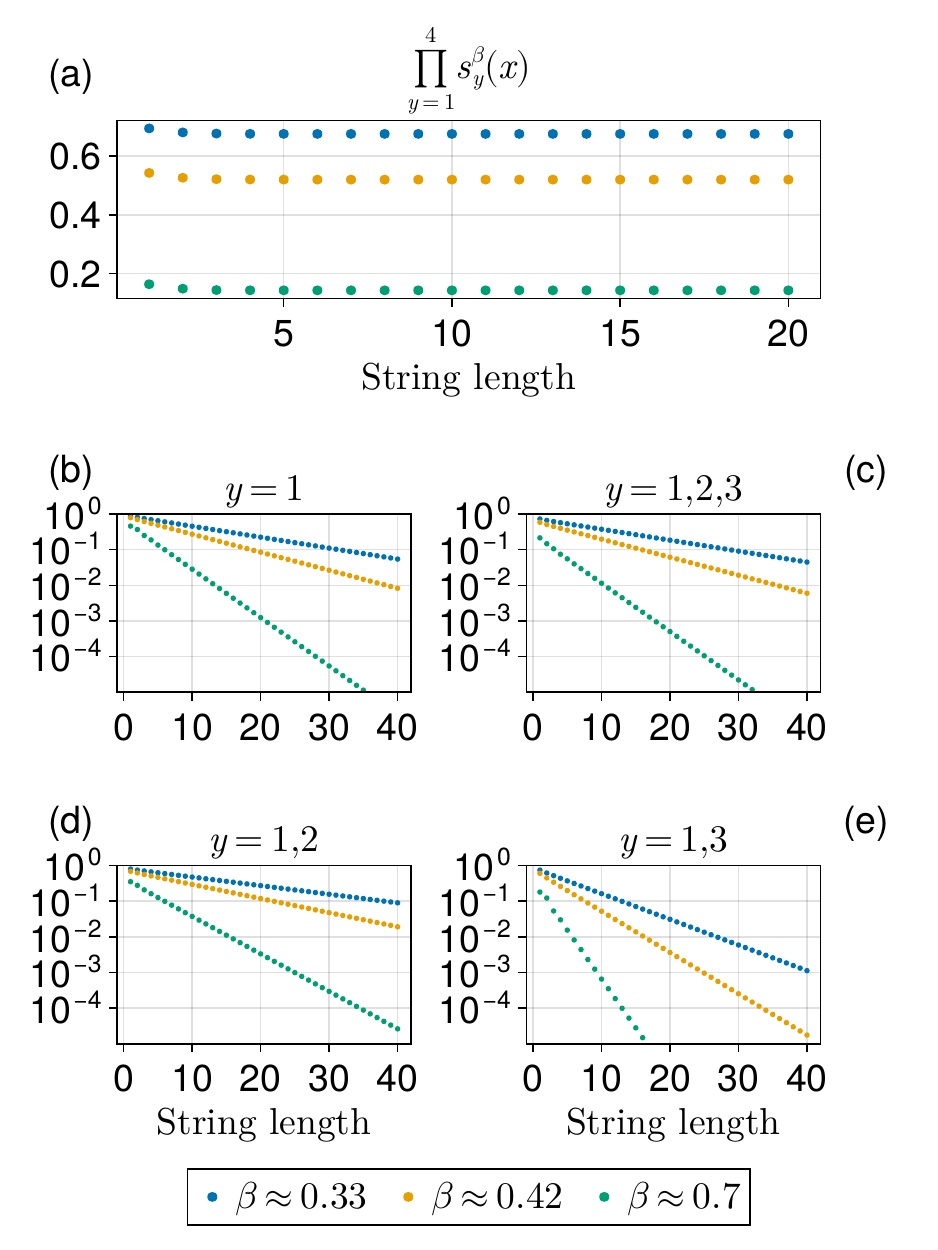}
    \caption{Expectation values for different string operators obtained from the infinite MPS of the XC-4(0) cylinder in the semion sector. The infinite MPS has bond dimension $\chi = 1000$. (a) The string operator goes through all the chains. This quantity saturates at a nonzero constant for $x\gtrsim 4$. (b--e) The string operator does not involve all four chains. The expectation value decays exponentially as a function of the length $x$.} 
    \label{FourChainFig2}
\end{figure}

Finally, we examine different string operators on infinite three- and four-chain cylinders. For three-chain cylinders, there are three different types of strings, namely the single string $s_1^\beta(x)$, a product of two strings $\prod_{y=1}^2 s_y^\beta(x)$, and the full string $\prod_{y=1}^3 s_y^\beta(x)$. Using the two different ground states with or without a topological flux, we find that only the full string develops a nonzero expectation value. The other two strings decay exponentially when the length of the strings increases (Fig.~\ref{ThreeChainString}). The same behavior can also be found using the infinite MPS of the XC-4(0) four-chain cylinder (Fig.~\ref{FourChainFig2}).

\subsection{Two-dimensional results}

To supplement the investigation on cylinders of finite circumference illustrated in the previous sections, we employ iPEPS \cite{verstraete2004PEPS, Jordan2008iPEPS} to study the behavior of the model described in Secs.~\ref{Coupled-wire} and ~\ref{SLL} in the two-dimensional thermodynamic limit.
The iPEPS is a subclass of tensor network states defined directly on infinite two-dimensional lattices. 
They fulfill by construction an area law of entanglement entropy and are thus well suited to the study of ground states of local Hamiltonians in two spatial dimensions. 
We note that, by contrast, one-dimensional ans\"atze like MPS do not fulfill the area law in two dimensions. Hence, if applied to pseudo-two-dimensional geometries like finite-circumference cylinders, one needs to use bond dimensions that are exponentially large in the cylinder circumference. 
Once equipped with state-of-the-art variational optimization~\cite{Corboz16Vari, Liao19ADiPEPS, Weerda24VariPEPS, PhysRevB.111.235116}, the utility of the iPEPS ansatz for the numerical study of chiral topological states has recently been demonstrated both in the case of spin systems~\cite{Hasik22CSL} as well as in the study of mobile bosons in a static gauge field~\cite{Weerda24FQH}.

\begin{figure}[hbt]
    \centering
    \includegraphics[width=\linewidth]{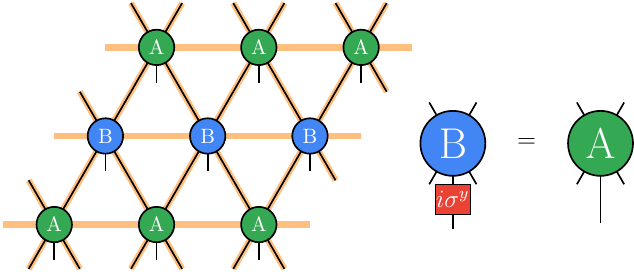}
    \caption{\textit{Left:} Illustration of the iPEPS ansatz in the triangular lattice. The straight black lines represent indices of the local tensors. The broad orange lines highlight the underlying triangular lattice. We do not explicitly enforce the global $U(1)$ symmetry in the tensor network. \textit{Right:} The relation of the different local tensors used to achieve a translation-invariant ansatz. We relate the two local tensors via a sub-lattice spin-rotation.}
    \label{fig:iPEPS_ansatz}
\end{figure}

\begin{figure}[tb]
    \centering
    \includegraphics[width=\columnwidth]{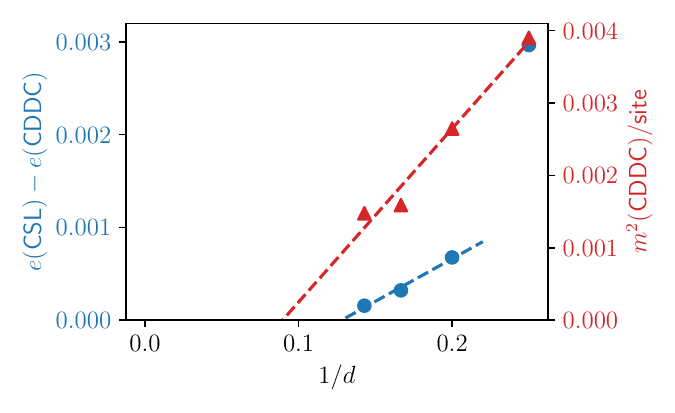}
    \caption{Extrapolations of the numerical results of the iPEPS simulations over the inverse bulk bond dimension $1/d$. \textit{Left $y$ axis (blue):} Difference of the energy density $e = E/N$ between the CSL state with the sublattice structure as shown in Fig.~\ref{fig:iPEPS_ansatz} and a $\begin{smallmatrix}C & D \\ D & C\end{smallmatrix}$ ansatz with two independent tensors $C$ and $D$. \textit{Right $y$ axis (red):} Staggered magnetization $m^2$ per site of the $\begin{smallmatrix}C & D \\ D & C\end{smallmatrix}$ ansatz.}
    \label{fig:ES_iPEPS_diff}
\end{figure}

Generically, the correlations of the system require an iPEPS ansatz consisting of an alternation between two independent local tensors on even and odd rungs of the model. The local tensors within one rung can be chosen to be identical. To be able to use a fully translation-invariant iPEPS ansatz, we perform a sub-lattice rotation $U_{\text{sub}} = \prod_{j} i\sigma^y_j$, where the product runs over all lattice sites on even rungs of the system. We can use this ansatz since the Hamiltonian has a $U(1)$ symmetry we can utilize~\cite{PhysRevLett.133.176502}.
We illustrate this ansatz in Fig.~\ref{fig:iPEPS_ansatz}.
With this ansatz, we find ground states for several bond dimensions $d$, whose energy density $e$ and local magnetization $m^2$ we list in Table~\ref{Tab.iPEPS}, and we further analyze the states via their entanglement spectrum below. We want to note that we found in the iPEPS simulation a competing ground state that breaks the sublattice structure and needs a $\begin{smallmatrix}C & D \\ D & C\end{smallmatrix}$ unit cell with two independent tensors $C$ and $D$. For the maximal available bond dimension up to $d=7$, this state is slightly lower in energy.
With available data, we could not resolve if this is an effect of the finite bond dimension and the additional restriction due to the sublattice ansatz.
But a extrapolation of the energy difference between both states as shown in Fig.~\ref{fig:ES_iPEPS_diff} (left/blue $y$ axis) suggests that the states will coincide in the limit of larger dimensions.
Additionally, the magnetization of the $\begin{smallmatrix}C & D \\ D & C\end{smallmatrix}$ state seems to vanish in the extrapolation to large bond dimensions as well as shown in Fig.~\ref{fig:ES_iPEPS_diff} (right/red $y$ axis).

\begin{table}[thb]
\caption{\label{Tab.iPEPS} Energy densities $e = E/N$ and local magnetizations $m^2$ of the variational iPEPS states at different bond dimensions.}
{\begin{tabular}{ ccccc } 
 \hline\hline
 bond dim. & $e$(CSL) & $m^2$(CSL) & $e$(CDDC)  & $m^2$(CDDC)\\ 
  \hline 
 $d = 4$ & $-0.32624$ & $7.22\times 10^{-5}$ & $-0.32921$ & $3.90\times 10^{-3}$ \\ 
 $d = 5$ & $-0.33031$ & $3.41\times 10^{-6}$ & $-0.33098$ & $2.65\times 10^{-3}$ \\ 
 $d = 6$ & $-0.33166$ & $5.55\times 10^{-7}$ & $-0.33198$ & $1.59\times 10^{-3}$ \\
 $d = 7$ & $-0.33208$ & $1.30\times 10^{-6}$ & $-0.33224$ & $1.48\times 10^{-3}$ \\
 \hline\hline
\end{tabular}}
\end{table}

\begin{figure}[tb]
    \centering
    \includegraphics[width=\linewidth]{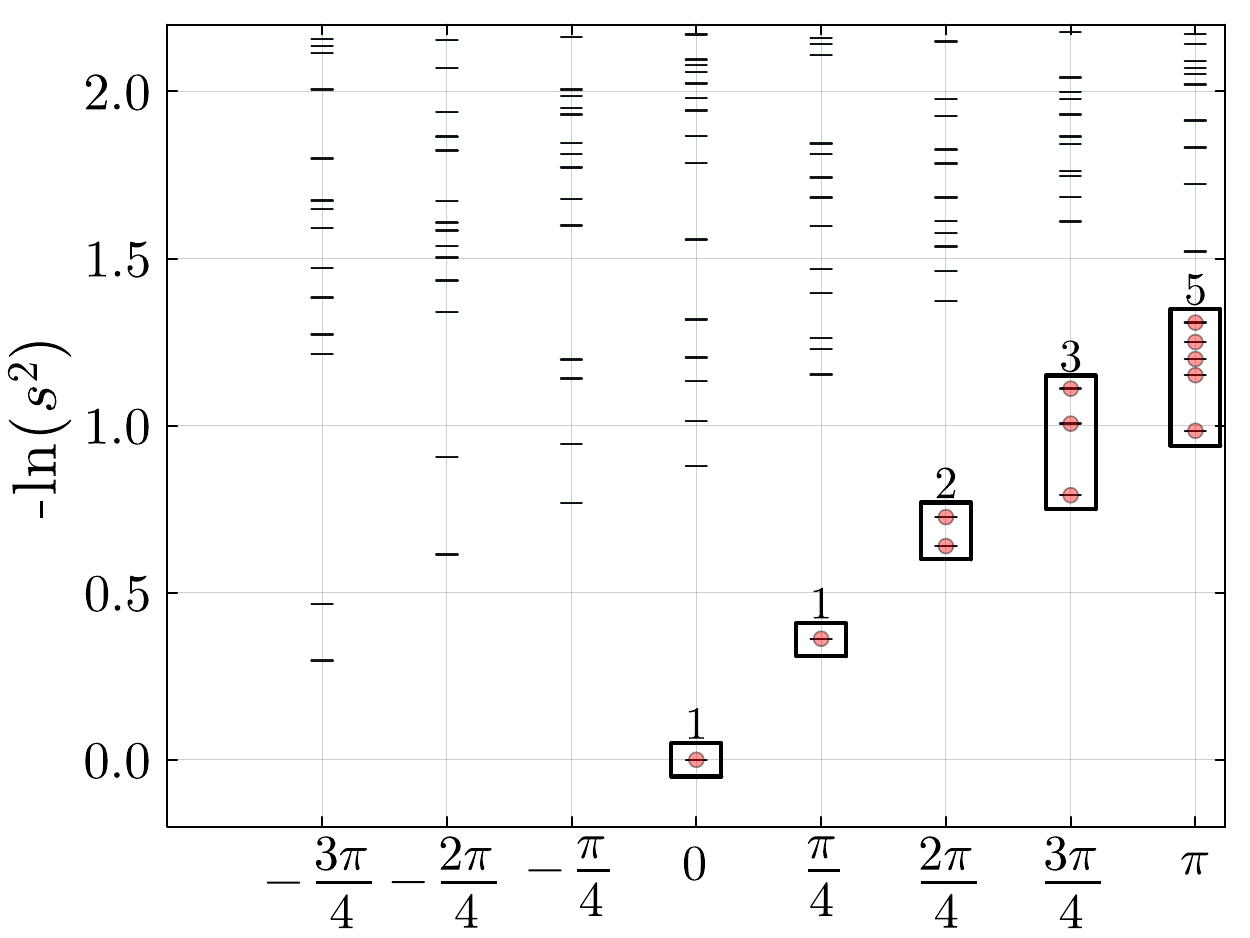}
    \caption{Entanglement spectrum corresponding to the low-lying edge-spectrum obtained via PEPS bulk-boundary correspondence. The discrete momenta are due to the circumference-eight cylinder on which the bipartition is considered. We clearly find the degeneracy counting associated with the partition of integers $(1,1,2,3,5, \cdots)$, as is expected for the Kalmayer-Laughlin chiral spin liquid.  Additionally, a double degenerate branch of the spectrum is noticeable at negative momenta~\cite{Cincio13PRL, ArildsenPRB22}. As we are optimizing the iPEPS on a plane, we only access the identity sector in this calculation.}
    \label{fig:ES_iPEPS}
\end{figure}

Once we have obtained an approximation of the ground state in terms of an iPEPS wavefunction, we can access information about its edge spectrum by employing the PEPS bulk-boundary correspondence~\cite{Cirac11PEPSBBCorresp}.
This allows us to compute the entanglement spectrum of a bipartition of the system, which, following Li and Haldane~\cite{Li08PRL}, is related to the low-lying physical spectrum on the edge. To observe the degeneracy counting of the chiral entanglement spectrum, we are setting our system on a finite cylinder to discretize the momenta. We show the entanglement spectrum obtained on a circumference-eight cylinder in Fig.~\ref{fig:ES_iPEPS}.
We can clearly identify a chiral spectrum and the degeneracy counting of the partition of integers associated with the edge spectrum of the Kalmeyer-Laughlin spin liquid.

\section{Discussion} \label{Discussion}

In this article, we have demonstrated how to obtain a microscopic lattice Hamiltonian of the Kalmeyer-Laughlin CSL from a coupled-wire construction.
The crucial step lies in introducing nonperturbative intrawire interactions to realize a gapless fixed point (sliding Luttinger liquid, SLL) whose strongest instability is toward the CSL.
We constructed these couplings with the help of a duality that allowed us analytical control over the parameters of the SLL. 

By using extensive numerical simulations of quasi-one-dimensional systems with DMRG and VUMPS and in two dimensions via iPEPS, we confirmed that the lattice model realizes a CSL ground state.
The connection between the lattice models and coupled-wire analysis allowed us to make sharp numerical predictions about the CSL beyond spectral flow and entanglement spectra. In particular, we showed that infinite cylinders of different circumferences $N$ follow a specific sequence of quasi-one-dimensional phases. Additionally, we introduced nonlocal operators that serve as order parameters for numerical simulations. 

The lattice models that we study include multispin interactions, which are sometimes regarded as unnatural. However, we note that all the interactions in our models arise in the $t/U$ expansion of a generalized Hubbard model with the appropriate symmetries \cite{Macdonald88PRB, Garuchva25JPA, makuta2025effectivespinmodelanisotropic}. Consequently, one may expect such Hamiltonians to arise for ``weak'' Mott insulators, where virtual charge fluctuations are significant. Unfortunately, the precise relation between couplings used in this study is not achievable in a nearest-neighbor model with purely on-site repulsion. Still, longer-range hoppings or interactions could improve this situation and better approximate the spin models we investigated here. Alternatively, quantum simulators may be able to directly engineer the spin-spin interactions.


This work demonstrates a prescription to turn coupled-wire constructions from analytical toy models to concrete microscopic systems. 
All the analytical and numerical steps in our analysis can be adapted for known coupled-wire constructions of other spin liquids and fractional quantum Hall or Chern insulators~\cite{Kane02prl, Teo14PRB, Patel16PRB, Huang16PRB, Huang17PRB, Lecheminant17PRB, Chen19PRB, leviatan2020, leviatan2022, Shavit24PRL}. 
In particular, the duality relation used to construct suitable SLL fixed points readily generalizes to those contexts: the operator $S^z$ in Eq.~\eqref{eqn.SzRotation} is replaced by the particle number in itinerant models. 

For more exotic topological states, coupled wire constructions use multichannel Luttinger liquids as a starting point. Reference \cite{Teo14PRB} constructed wire models of $SU(2)_n$ topological orders based on $n$-channel Luttinger liquids.
In those cases, the single analytically controlled parameter $\alpha$ in the SLL of Eq.~\ref{eqn.Fixedpoint} can be replaced by $n^2$ parameters $\alpha_{\sigma,\sigma'}$. 
The corresponding lattice model can be obtained by reversing the analysis of Sec.~\ref{SLL} with a generalized duality transformation. 
Specifically, Eq.~\eqref{Duality} can be generalized to dress $S^-_{\vect r,\sigma}$ with independent string operators for each channel $\sigma'$ on the neighboring wires. 
We expect that this level of analytical control over the long-wavelength properties of a lattice model will permit the realization of more exotic topological phases, including non-Abelian states.

Numerical studies of spin-1 and spin-3/2 on the triangular lattice showed evidence of CSLs with non-Abelian $SU(2)_2$ and $SU(2)_3$ order, respectively \cite{Huang22PRB, Luo23PRB}. Together with the Kalmeyer-Laughlin CSL in spin-1/2 system, these findings suggest a pattern of $SU(2)_{2S+1}$ spin liquids for spin $S$. Comparison with the coupled-wire construction mentioned in the previous paragraph implies that this spin corresponds to $n=2S+1$ channels. To understand this relation, we note that spin $S$ models can generally realize an $SU(2S+1)_1$ Wess-Zumino-Witten \cite{WessZumino, Witten} conformal field theory, which has central charge $c=2S+1$. (For example, the biquadratic spin-1 chain at the Uimin-Lai-Sutherland point realizes the $SU(3)_1$ theory \cite{Uimin70JETP, Lai74JMP, Sutherland75PRB, Itoi97PRB}.) We expect that spin $S$ chains realizing these critical points would be suitable building blocks for constructing lattice models of $SU(2)_{2S+1}$ CSL.

\begin{acknowledgments}
The authors are indebted to Simon Trebst, Guo-Yi Zhu, Bo Han, and Yuval Gefen for illuminating discussions. We acknowledge support from the Deutsche Forschungsgemeinschaft (DFG) project Grant No. 277101999 within the CRC network TR 183 (Subproject No. A04), supported by a research grant from the Estate of Hermine Miller, the Sheba Foundation, and Dweck Philanthropies, Inc. T.G. gratefully acknowledges Luis E. F. Foa Torres for his hospitality in Santiago de Chile.
N.T. and M.R. are further supported by the DFG under Germany’s Excellence Strategy---Cluster of Excellence Matter and Light for Quantum Computing (ML4Q) EXC No. 2004\slash1390534769. E.L.W. thanks the Studienstiftung des deutschen Volkes for support. D.F.M. was
supported by the Israel Science Foundation (ISF) under Grant
No. 2572\slash21 and by the Minerva Foundation with funding
from the Federal German Ministry for Education and Research. Additionally, the authors gratefully acknowledge the
Gauss Centre for Supercomputing e.V. (\href{www.gauss-centre.eu}{www.gauss-centre.eu}) for funding this project by providing computing time through the John von Neumann Institute for Computing (NIC) on the GCS Supercomputer JUWELS \cite{JUWELS} at the J\"{u}lich Supercomputing Centre (JSC) (Grant No. NeTeNeSyQuMa) and the FZ J\"{u}lich for JURECA \cite{JURECA2021} (Institute Project No. PGI-8). 

\textit{Data Availability.} Most of the data that support the findings of this article are openly available \cite{gao_2025_17697831}. We skip uploading some tensor network states, such as MPSs involved in spin pumping simulations for five and six wires, due to the 50 GB quota of datasets in Zenodo. These data are available upon reasonable request from the authors.
\end{acknowledgments}

\appendix
\section{Duality transformation}\label{app.duality}

In Eq.~\eqref{Duality}, we introduced a transformation from $S^{x,y,z}$ to new variables $\tilde{S}^{x,y,z}$ that preserves their commutation relations. We then argued that Eq.~\eqref{DualityField} is the corresponding transformation from the field theory variables $\theta,\varphi$ to $\tilde \theta,\tilde \varphi$. Here, we provide a detailed derivation and numerical support.

\subsection{String operators}
The central element of this mapping is the string operator
\begin{align}
\label{eqn.app.sigma}
\sigma_{x,x+n} \equiv 2 \sum_{m = 0}^n S^z_{\bm{r} + m \bm{b}_0} = 2\sum_{m = 0}^n \tilde{S}^z_{\bm{r} + m \bm{b}_0}~,
\end{align}
which is a sum of commuting terms that each square to unity. 
We are suppressing the wire index $y$ when there is no danger of ambiguity to lighten the notation. The string operator enters the lattice duality transformation through
\begin{equation}\label{eqn.app.s}
   s^\alpha_{x,x+n}\equiv \prod_{j = 0}^n R^\alpha_{\bm{r} +j\bm{b}_0}= \exp{-i \alpha \frac{\pi}{2} \sigma_{x,x+n}}~,
\end{equation}
which satisfies the periodic property
\begin{align}\label{eqn.app.speriod}
    s^{\alpha+2}_{x,x+n}  =  (-1)^{n+1} s^{\alpha}_{x,x+n}~.
\end{align}

To obtain expressions of $\sigma$ and $s$ in terms of $\theta,\varphi$, we use Eq.~\eqref{eqn.Sz} in Eq.~\eqref{eqn.app.sigma} and drop the oscillatory term to obtain 
\begin{align}
\sigma_{x,x+n} \approx \frac{2}{\pi}\left[\theta(x+n)-\theta(x) \right] =\frac{2}{\pi}\left[\tilde\theta(x+n)-\tilde\theta(x) \right].
\end{align}
Inserting this relation into Eq.~\eqref{eqn.app.s} yields 
\begin{align}
\label{eqn.app.sfield}
   s^{\alpha}_{x,x+n} \approx e^{i \alpha [\theta(x+n)-\theta(x)]}~,
\end{align}
which does not satisfy Eq.~\eqref{eqn.app.speriod}. To enforce the periodic property, one could instead take
\begin{equation}
\label{eqn.app.sfieldperiodic}
\begin{split}
    s^{\alpha}_{x,x+n} &\sim \sum_{k} r_{0}^{(k)}e^{i \left(\alpha + 4k\right) \left[\theta\left(x+n\right)-\theta(x)\right]}\\
    &+ (-1)^n \sum_{k} r_{\pi}^{(k)} e^{-i \left(2-\alpha + 4k\right) \left[\theta\left(x+n\right)-\theta(x)\right]}~.
\end{split}
\end{equation}
However, unless $\alpha$ is an odd integer, there is a single component on the sum on the right-hand side that dominates at large $n$. In particular, for $\alpha \in (-1,1)$, the dominant contribution is the one in Eq.~\eqref{eqn.app.sfield}.

\begin{figure}[tb]
    \centering
    \includegraphics[width=\linewidth]{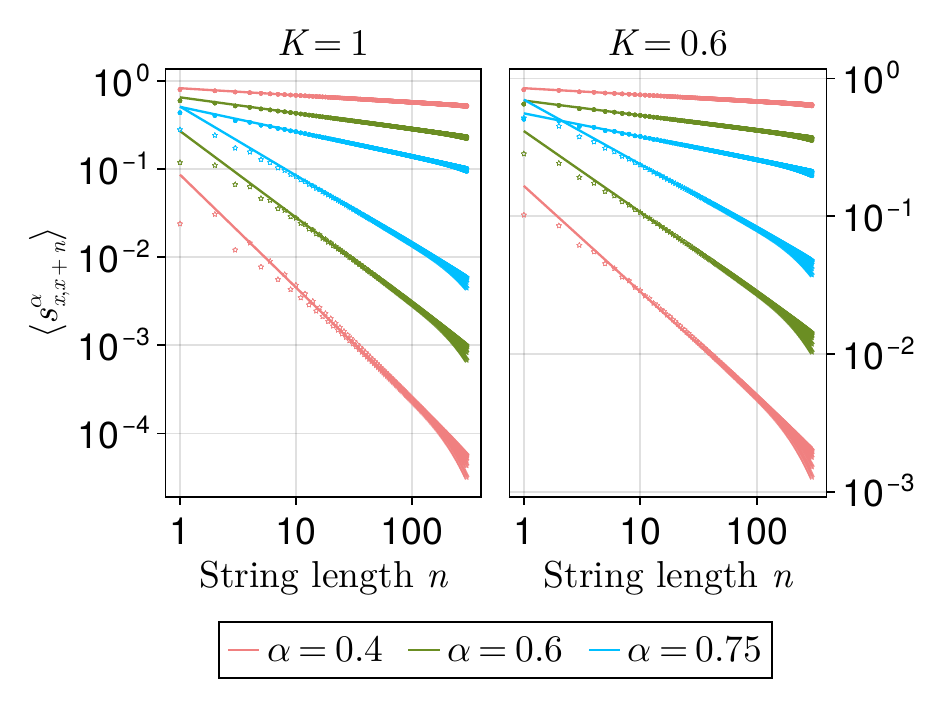}
    \caption{
    String operator expectation values of $XXZ$ chains with different parameter $K$. Data are obtained from VUMPS at bond dimensions $50, 60, \cdots, 100$ without conserving the total $S^z$. The (slower) branches with solid dots describe the uniform part in the expectation value, while the (faster) branches with hollow dots belong to the staggered part. We averaged over two sites to obtain the uniform part and used the second-order central difference to get the staggered part. Solid lines are fitted by the data at bond dimension 100 with fixed power-law exponents.}
    \label{StringExpectation}
\end{figure}

In a single chain with Luttinger parameter $K$, we find
\begin{align}
\label{eqn.app.uniformstagg}
   \expval{s^{\alpha}_{x,x+n}} \sim r_0 n^{-\alpha^2 K / 2}+ r_\pi (-1)^n n^{-(2-\alpha)^2 K / 2}+\cdots ~,
\end{align}
In Fig.~\ref{StringExpectation}, we show numerical data obtained from a single $XXZ$ chain using VUMPS up to bond dimension $100$. Both the uniform and the staggered parts of the expectation value follow the expected power laws.

\subsection{Transformations on the field variables}
Since the spin $z$ components are untouched during the nonlocal transformation Eq.~\eqref{Duality}, the fields $\theta_y$ and $\tilde{\theta}_y$ are identical, i.e., $\theta_y =\tilde{\theta}_y$. To infer how $\varphi_y$ is related to the dual variables, we note that the dual spin operators $\tilde{S}^-_{\vect{r}}$ are defined in Eq.~\eqref{Duality} as
\begin{align}
    \tilde {S}^-_{\bm{r}} = S^-_{\bm{r}} \cdot s^\alpha_{-\infty,\bm{r} - \bm{b}_-} \cdot s^\alpha_{\bm{r} + \bm{b}_-,+\infty}~.
\end{align}
Using Eq.~\eqref{eqn.app.sfield} and dropping terms at infinities, we find
\begin{align}
    \tilde {S}^-_{x,y} \approx {S}^-_{x,y}  \cdot e^{i \alpha \theta_{y-1}(x)}\cdot e^{-i \alpha \theta_{y+1}(x)}~.
\end{align}
Since $\tilde S^- \propto e^{i \tilde\varphi}$ and $S^- \propto e^{i \varphi}$, we conclude that $\varphi_y \equiv \Tilde{\varphi}_y - \alpha \Tilde{\theta}_{y-1} + \alpha \Tilde{\theta}_{y+1}$.

\subsection{Numerical tests of the duality}

\begin{figure}[tb]
    \centering
    \includegraphics[width=\linewidth]{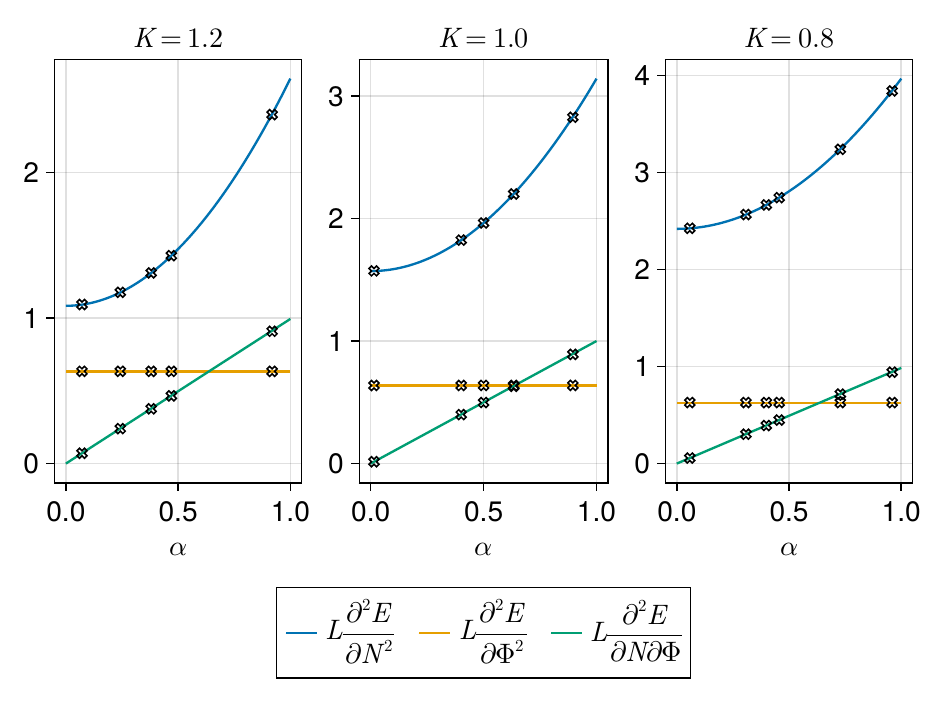}
    \caption{Finite-difference approximations to different second derivatives of the ground-state energy. The difference step in $N$ is $\delta N = 2$, and the step in $\Phi$ is $\delta \Phi = \pi/50$. The solid lines correspond to the right-hand side of Eq.~\eqref{eqn.thermal}, and the crosses are data obtained from DMRG. We simulated a periodic ladder with $2 \times 32$ sites and kept $3000$ states during the last DMRG iteration.}
    \label{fig.LLparamsDuality}
\end{figure}

We tested the validity of Eq.~\eqref{eqn.Fixedpoint} on a two-leg ladder by numerically computing thermodynamic quantities that are fully determined by the field theory parameters $u,K,\alpha$. Specifically, we used periodic boundary conditions along $x$ with opposite twist $\pm \Phi$ for the two chains and open boundaries in the $y$ direction. The lattice Hamiltonian for this system is the same as what was studied in Ref.~\cite{Pozsgay22PRE}. In this case, the field-theoretic duality mapping reduces to $\varphi_1 = \tilde \varphi_1 +\alpha \tilde{\theta}_2$ and $\varphi_2 = \tilde \varphi_2 -\alpha \tilde{\theta}_1$. We thus obtain the long-wavelength Hamiltonian density
\begin{equation}\label{eqn.twochainfield}
\begin{split}
\cH &= \frac{u}{2\pi} \sum_{y=1}^2 \left[\left(\frac{1}{K} + \alpha^2 K\right) \left(\grad \theta_y \right)^2 + K \left(\grad \varphi_y\right)^2\right]\\
&+ \frac{u K}{\pi} \alpha \left[\left(\grad \varphi_1\right) \left(\grad \theta_2 \right) - \left(\grad \varphi_2\right) \left(\grad \theta_1 \right)\right].
\end{split}
\end{equation}
It is straightforward to compute the ground-state energy $E$ of the quadratic theory as a function of the twist $\Phi$, system size $L$, and the total spin $N\equiv-\frac{1}{\pi} \int_x \, \left(\partial_x \theta_1 + \partial_x \theta_2\right)$. In particular, we find the derivatives evaluated at $\Phi=N=0$ to be
\begin{equation}\label{eqn.thermal}
\begin{split}
    L\frac{\partial^2 E}{\partial N^2} &= \frac{\pi}{2} \left(\frac{u}{K} + \alpha^2 u K\right)~,\\
    L\frac{\partial^2 E}{\partial \Phi^2} &= \frac{2}{\pi} u K~,\\
    L\frac{\partial^2 E}{\partial N \partial \Phi} &= u K \alpha~.
\end{split}
\end{equation}
Our numerical DMRG simulations on a $2 \times 32$ system are shown in Fig.~\ref{fig.LLparamsDuality} and demonstrate excellent agreement with the analytical prediction.

\subsection{Duality near odd integer $\alpha$}

When deriving the duality transformation of the long-wavelength variables, we kept only the leading term in Eq.~\eqref{eqn.app.sfieldperiodic}. By comparing the two terms in Eq.~\eqref{eqn.app.uniformstagg}, we estimate a length scale $\ell$ beyond which this approximation is valid. Specifically, we require that
\begin{align}
 \epsilon \equiv \frac{\ell^{-(2-\alpha)^2 \frac{K}{2}}}{\ell^{-\alpha^2 \frac{K}{2}}}  \ll 1~.
\end{align}
As $\alpha \rightarrow 1$, the length scale $\ell$ diverges, and there is no parametric regime where dropping the subleading term would be justified. 

We note that $\alpha \rightarrow \pm 1$ yield the same microscopic models, but the scaling dimensions of ${\cal O}_\text{chiral}$, according to Eq.~\eqref{eqn.deltachiral}, differ significantly in the two limits. Reassuringly, the time reversed partner of Eq.~\eqref{eqn.antichiral}, i.e., 
\begin{equation}
{\cal O}'_{\text{chiral}} = \cos{\left(2\theta_{y-1} + 2\theta_{y+2} + \varphi_y - \varphi_{y+1}\right)}~,
\end{equation}
which realizes the same topological phase as ${\cal O}_\text{chiral}$ has a scaling dimension that satisfies $\Delta_{\text{antichiral}}\big|_{-2-\alpha} = \Delta_{\text{chiral}}'\big|_{\alpha}$.

\section{Vertical spin exchange and its low-energy description}\label{app.psll}

We utilize Eq.~\eqref{eqn.Sz} to identify the most relevant operator out of the lattice-scale inter-chain $S^z S^z$ interaction
\begin{align}
    H_\rho = J^z_\perp \sum_{i = +, -} \sum_{\bm{r}} \left(S^z_{\bm{r}} S^z_{\bm{r} + \bm{b}_{i}} + \text{H.c.}\right)~.
\end{align}
Contributions from the staggered component in Eq.~\eqref{eqn.Sz} are suppressed due to the triangular lattice structure. Explicitly, we identify
\begin{align}
    {\cal H}_\rho \sim g_\rho \sum_{y}\left(\partial_x \theta_y\right) \left(\partial_x \theta_{y+1}\right) + \cdots~,
\end{align}
where we have neglected less relevant operators and the coupling strength
\begin{equation}\label{eqn.app.perturbation}
    g_\rho = \frac{2J^z_\perp}{\pi^2} \equiv \frac{u \lambda}{\pi K}
\end{equation}
when $J^z_\perp \ll 1$. 

Beyond the perturbative regime, the approximation Eq.~\eqref{eqn.app.perturbation} is tested numerically on an open-boundary four-chain cylinder. The total magnetization $N_y$ on each chain is related to the field $\theta_y$ by
\begin{align}
    N_y = -\frac{1}{\pi} \int_x \expval{\partial_x \theta_y}~.
\end{align}
We consider three situations where
\begin{equation}
    \begin{cases}
        N_y = N, & \text{Case 1},\\
        N_y = (-1)^y N, & \text{Case 2},\\
        N_1 = N, \quad N_2 = 0, \quad N_3 = N, \quad N_4 = 0, & \text{Case 3}.\\
    \end{cases}
\end{equation}
The corresponding second derivatives of the ground state energy are
\begin{equation}
    L\frac{\partial^2 E}{\partial N^2} =
    \begin{cases}
        4 \pi \frac{u}{K} \left(1 + 2 \lambda\right), & \text{Case 1},\\
        4 \pi \frac{u}{K} \left(1 - 2 \lambda\right), & \text{Case 2},\\
        2 \pi \frac{u}{K}, & \text{Case 3}.\\
    \end{cases}
\end{equation}
Combining the third equation with the first one or the second one, we can obtain $\lambda$ numerically through finite differences and denote the values as $\lambda_{13}$ and $\lambda_{23}$.

\begin{figure}[tb]
    \includegraphics[width=\columnwidth]{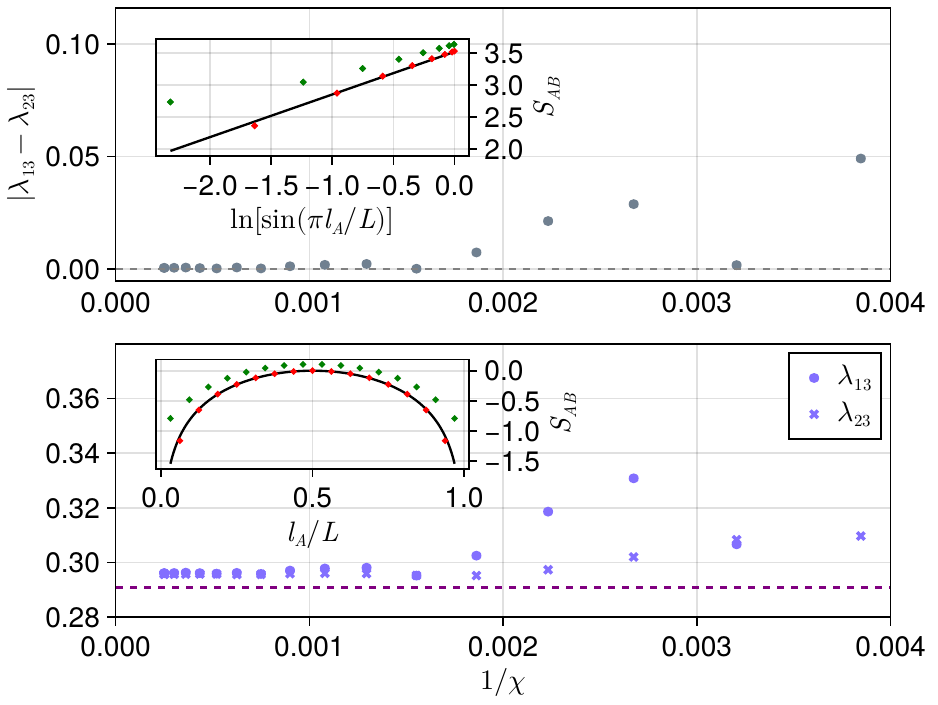}
    \caption{
    Numerical results of $\lambda_{13}$ and $\lambda_{23}$. Upper figure: $|\lambda_{13} - \lambda_{23}|$ versus $1/\chi$, where $\chi$ is the maximal bond dimension of the MPS. Lower figure: $\lambda_{13}$ and $\lambda_{23}$ versus $1/\chi$. The purple dashed line represents the perturbative theoretical value of $\lambda$.
    Inset figures: Ground-state entanglement entropy at $\chi = 4000$. The red (green) points represent a virtual cut that separates the system into two even(odd)-length subsystems $A$ and $B$. The black curves are the theoretical entanglement entropy of Ref.~\cite{Calabrese_2009}, assuming a central charge $4$.} 
    \label{FourChainJustification}
\end{figure}

The numerical results of $\lambda_{13}$ and $\lambda_{23}$ are shown in Fig.~\ref{FourChainJustification}. We obtain data up to bond dimension $4000$ from a $32$-unit-cell open XC-4(0) cylinder. At each given bond dimension $\chi$, we perform $10$ DMRG sweeps or obtain energies to a relative precision $10^{-5}$. The upper figure shows the difference between $\lambda_{13}$ and $\lambda_{23}$, which quantifies how much the low-energy effective theory deviates from Eq.~\eqref{eqn.Fixedpoint}.
Furthermore, we compare the two values with the perturbative expression 
\begin{equation}\label{eqn.app.perturbative}
    \lambda = J^z_\perp \frac{K}{u} \frac{2}{\pi}
\end{equation}
in the lower figure. The small deviations of the values with respect to Eq.~\eqref{eqn.app.perturbative} allow us to extend this relation moderately beyond the perturbative regime.

\section{Scaling dimensions}\label{app.scalings}

The SLL fixed point in Eq.~\eqref{eqn.Fixedpoint}, together with all symmetry-allowed perturbations, captures the low-energy physics of our lattice model as long as the interwire couplings $J^z_{\perp}$ and $J_{\perp}$ are sufficiently small. When the perturbation $\cO_{\text{chiral}}$ has the smallest scaling dimension, the system will flow to the CSL phase. To quantify this statement, we deduce the scaling dimensions of 
\begin{align}
    \cO_{\text{chiral}} &= \cos{\left(2\theta_y + 2\theta_{y+1} + \varphi_y - \varphi_{y+1}\right)}~, \notag \\
    {\cal O}'_{\text{antichiral}} &=  \cos{\left(2\theta_{y-1} + 2\theta_{y+2} - \varphi_y + \varphi_{y+1}\right)}~, \notag \\
    \cO_{XY} &=  \cos{\left(\varphi_y - \varphi_{y+2}\right)}~, \notag\\
    \cO_{\text{spiral}} &=  \left(\grad \varphi_y + \grad \varphi_{y+1}\right) \sin{\left(\varphi_y - \varphi_{y+1}\right)}~, \notag\\
    \cO^{\pm}_{\theta} &=  \cos{\left(2\theta_y \pm 2\theta_{y+2}\right)}~.
\end{align}
When the operators $\cO^{+}_{\theta}$ and $\cO^{-}_{\theta}$ flow to strong coupling, they give the same effect as $\cO_{\text{VBS}}$ or $\cO_{\text{Ising}}$, depending on how the arguments are pinned to extrema. Besides, ${\cal O}'_{\text{antichiral}}$ defined here can also lead to the CSL with an opposite chirality similar to ${\cal O}_{\text{antichiral}}$ in the main text, whose scaling dimension is generally larger at a nonzero positive $\alpha$.

We obtain the scaling dimensions using the dual formulation in terms of the field variables $\Tilde{\varphi}_y$ and $\Tilde{\theta}_{y'}$. Correlation functions involving $\Tilde{\varphi}_y$ and $\Tilde{\theta}_{y'}$ take relatively simple forms at the SLL of Eq.~\eqref{eqn.Fixedpoint}. The fixed point is translationally invariant in the transverse direction; therefore, we introduce Fourier modes for the fields as
\begin{equation}
    \Tilde{\varphi}_y = \frac{1}{\sqrt{N}} \sum_{q} e^{i q y} \Tilde{\varphi}_q, \qquad \Tilde{\theta}_y = \frac{1}{\sqrt{N}} \sum_{q} e^{i q y} \Tilde{\theta}_q.
\end{equation}
Using these new variables, we rewrite Eq.~\eqref{eqn.Fixedpoint} as
\begin{equation}
    \cH_{\text{CSLL}} = \frac{u}{2\pi} \sum_q  \left(\frac{1 + 2\lambda \cos{q}}{K} |\partial_x \Tilde{\theta}_q|^2 + K |\partial_x \Tilde{\varphi}_q |^2\right)~.
\end{equation}
Modes at different momentum $q$ decouple. For each momentum $q$, the effective Luttinger parameter is 
\begin{align}
K_q(\lambda) = \frac{K}{\sqrt{1 + 2 \lambda \cos{q}}}~.
\end{align}
We conclude that the field correlation functions are
\begin{align}
\expval{|\Tilde{\theta}_q(r) - \Tilde{\theta}_q(0)|^2} &\sim K_q(\lambda) \log{|r|}~,\\
\expval{|\Tilde{\varphi}_q(r) - \Tilde{\varphi}_q(0)|^2} &\sim \frac{\log{|r|}}{K_q(\lambda)}~.
\end{align}
Correlation functions such as $\expval{\Tilde{\theta}_y(r) \Tilde{\theta}_{y'}(0)}$ can be expressed using these correlators. We find
\begin{equation}
    \begin{split}
    \expval{\Tilde{\theta}_y(r) \Tilde{\theta}_{y+n}(0)} &= \frac{1}{N} \sum_q K_q(\lambda) \cos{\left(nq\right)} \log{|r|}~.
    \label{eqn.app.finite}
    \end{split}
\end{equation}
In the two-dimensional limit where $N \rightarrow \infty$, the summation becomes an integral, and we can rewrite it as 
\begin{equation}
    \begin{split}
        \expval{\Tilde{\theta}_y(r) \Tilde{\theta}_{y+n}(0)} &=  \log{|r|} \frac{K}{2\pi} \int_0^{2\pi} \d \varphi \, \frac{\cos{\left(n\varphi\right)}}{\sqrt{1 + 2 \lambda \cos{\varphi}}}\\
        &\equiv K f^{-}_n(\lambda) \log{|r|}~,
    \end{split}
\end{equation}
where we introduced 
\begin{equation}
    f^{\pm}_n(\lambda) = \frac{1}{2\pi} \int_0^{2\pi} \d \varphi \, \cos{\left(n\varphi\right)} \left(1 + 2 \lambda \cos{\varphi}\right)^{\pm \frac{1}{2}}
\end{equation}
to lighten equations. Similarly, the expression for the other field operator correlation function $\expval{\Tilde{\varphi}_y(r) \Tilde{\varphi}_{y'}(0)}$ is 
\begin{equation}
    \begin{split}
    \expval{\Tilde{\varphi}_y(r) \Tilde{\varphi}_{y'}(0)} &= \frac{1}{N} \sum_q \frac{1}{K_q(\lambda)} \cos{\left(nq\right)} \log{|r|}\\
    &\xrightarrow{N \rightarrow \infty} \frac{\log{|r|}}{K} f^{+}_n(\lambda)~.
    \end{split}
\end{equation}
Since $\cos{(\varphi)}$ is bounded between $-1$ and $1$, we can expand $f^{\pm}_n(\lambda)$ as a power series when $\lambda < 1$, i.e.,
\begin{equation}\label{Expansion}
    \begin{split}
        f^{\pm}_{2k}(\lambda) 
        &= \sum_{m=k}^\infty \lambda^{2m} \binom{\pm \frac{1}{2}}{2m} \frac{(2 m)!}{(m-k)! (m+k)!}~,\\
        f^{\pm}_{2k+1}(\lambda) &= \sum_{m=k}^\infty \lambda^{2m+1} \binom{\pm \frac{1}{2}}{2m+1} \frac{(2 m + 1)!}{(m-k)! (m+k+1)!}~.
    \end{split}
\end{equation}
In particular, $f_n^\pm(\lambda)= O(\lambda^n)$ and $f_0$ do not receive corrections at the first order in $\lambda$.

All scaling dimensions in the two-dimensional limit ($N \rightarrow \infty$) can be expressed as a linear combination of $f^\pm_n(\lambda)$ and thus inherit a power series expansion from Eq.~\eqref{Expansion}. 
One needs, at most, six field operator correlation functions to get the correlation function of the CSL operator $\cO_{\text{chiral}}$. From its correlation function, we obtain its scaling dimension as 
\begin{equation}
    \begin{split}
        \Delta_{\text{chiral}} &= K \left[\frac{(2-\alpha)^2 + \alpha^2}{2} f_{0}^-(\lambda) + \frac{4 - \alpha^2}{2} f_{1}^-(\lambda)\right]\\
        &+ K \left[\left(2 - \alpha\right) \alpha f_{2}^-(\lambda) + \frac{\alpha^2}{2} f_{3}^-(\lambda)\right]\\
        &+ \frac{1}{2K} \left[f_{0}^+(\lambda) - f_{1}^+(\lambda)\right].
    \end{split}
\end{equation}
The other scaling dimensions can be obtained in the same manner. The operator leading to the CSL with an opposite chirality has a scaling dimension $\Delta'_{\text{antichiral}}(\alpha) = \Delta_{\text{chiral}}(2-\alpha)$. The results for the other four operators are
\begin{align}
    &\Delta_{\theta}^{\pm} = 2K \left[f_{0}^-(\lambda) \pm f_{2}^-(\lambda)\right]~,\\
    &\Delta_{\text{spiral}} = 1 + \frac{1}{2K} \left[f_{0}^+(\lambda) - f_{1}^+(\lambda)\right]\notag\\
    &+ K \left[ \alpha^2 f_{0}^-(\lambda) - \frac{\alpha^2}{2} f_{1}^-(\lambda) - \alpha^2 f_{2}^-(\lambda) + \frac{\alpha^2}{2} f_{3}^-(\lambda)\right]~,\\
    &\Delta_{XY} = K \left[ \frac{3}{2} \alpha^2 f_{0}^-(\lambda) - 2 \alpha^2 f_{2}^-(\lambda) + \frac{\alpha^2}{2} f_4^-(\lambda)\right]\notag\\
    &+ \frac{1}{2K} \left[f_{0}^+(\lambda) - f_{2}^+(\lambda)\right]~.
\end{align}
These dimensions are plotted for $J^z_{\parallel} = -0.1$ and $\alpha = 0.75$ as functions of $\lambda$ in Fig.~\ref{ScalingD}. The figure also contains several plots for quasi-one-dimensional cylinders with finite $N$, for which Eq.~\eqref{eqn.app.finite} must be evaluated as a sum. We do not provide an explicit formula for arbitrary $N$, but the computation for any particular choice contains a finite number of terms and is thus straightforward.

\section{Transfer matrix} \label{app.tensors}

A wavefunction in the VUMPS algorithm is an infinite uniform MPS that is automatically translational invariant. For simplicity, we consider infinite MPSs invariant under one-site translation. Consequently, they are represented by a single tensor $A$, i.e., 
\begin{equation}
\ket{\Psi(A)}_{\text{VUMPS}} = 
\begin{tikzpicture}[baseline={([yshift=-.5ex]current bounding box.center)}]
    \Vertex[size = 0.5, x=1.5, label=$s_i$]{s2}

    \node at (2.5,0.5) {\ldots};
    \node at (0.5,0.5) {\ldots};

    \Vertex[shape = rectangle,x=0.5,y=1,style={color=white}]{A2l} 
    \Vertex[shape = rectangle,x=1.5,y=1, label=$A$]{A2}
    \Vertex[shape = rectangle,x=2.5,y=1,style={color=white}]{A2r}
    
    \Edge(A2)(s2)
    \Edge(A2l)(A2)
    \Edge(A2)(A2r)
    
\end{tikzpicture}~.
\end{equation}
Generalizations to multisite systems can be found, for example, in Ref.~\cite{Zauner18PRB}.

The dominant contribution to an inner product of two infinite MPS is encoded in the (mixed) transfer matrix. The overlap $\braket{\Psi(A)}{\Psi(B)}$ involves contractions of the mixed transfer matrix
\begin{equation}
T(A,B) = 
\begin{tikzpicture}[baseline={([yshift=-.5ex]current bounding box.center)}]
    \Vertex[shape = rectangle, y=1,style={color=white}]{ARl}
    \Vertex[shape = rectangle, style={color=white}]{ARbl}
    \Vertex[shape = rectangle, x=1, label=$B^\dagger$]{ARb}
    \Vertex[shape = rectangle,x=1,y=1, label=$A$]{AR} 
    \Vertex[shape = rectangle,x=2,y=1, style={color=white}]{AR2}
    \Vertex[shape = rectangle,x=2, style={color=white}]{AR2b}

    \Edge(ARl)(AR)
    \Edge(ARb)(AR)
    \Edge(ARbl)(ARb)
    \Edge(AR)(AR2)
    \Edge(ARb)(AR2b)
    
\end{tikzpicture}
\end{equation}
to an infinite number of times. A properly normalized infinite MPS $\ket{\Psi(A)}$, thus, should have $1$ as the largest eigenvalue of $T(A,A)$.

Another measure of the similarity between two infinite MPSs is the fidelity per site, which we have mentioned in Sec~\ref{Numerics}. Since the inner product between two infinite MPSs involves an infinite product of the mixed transfer matrix, even locally very similar states can be orthogonal to each other in the thermodynamic limit, given that the largest eigenvalue of $T(A,B)$ is less than $1$. Consequently, it is conventional to use the number instead, also named as the fidelity per site, to quantify how similar the two states are in the bulk of a system, i.e.,
\begin{equation}
    f(A,B) = \max_{T(A,B) v = \lambda v} |\lambda|~.
\end{equation}

Finally, the transfer matrix $T(A,A)$ can also reveal the correlation length of an infinite MPS $\ket{\Psi(A)}$. In our normalization scheme, the spectrum of the transfer matrix associated with a generic infinite MPS always has $1$ as the unique dominant eigenvalue \cite{Vanderstraeten19SciPost}. The second largest eigenvalue then determines the long-distance behaviors of the system. Consequently, the correlation length is
\begin{equation} \label{CorrelationLength}
\xi \equiv \max_{\lambda \neq 1} \frac{-1}{\ln{|\lambda|}}~.
\end{equation}

\section{Additional numerical data} \label{app.numerics}

In the main text, we have shown extensive numerical evidence in support of the CSL. Here, we provide additional data showing correlation lengths, entanglement spectra, and dependence on the bond dimension.

\subsection{Many-body energy gap and magnetic orders}

\begin{figure}[tb]
    \centering
    \includegraphics[width=\linewidth]{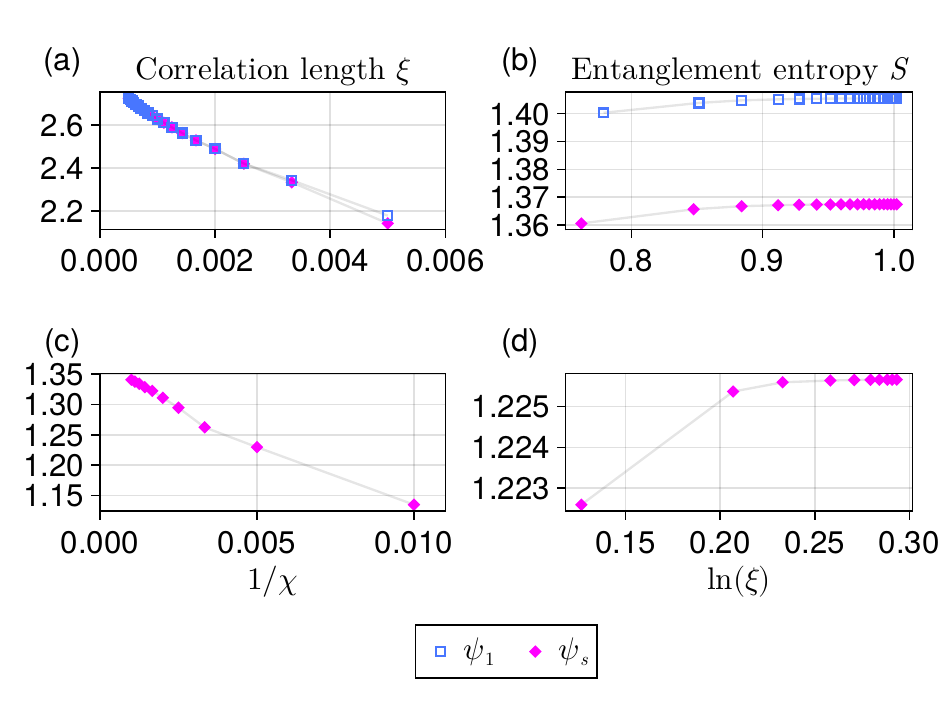}
    \caption{Correlation lengths and entanglement entropy for the XC-3 and XC-4(0) cylinders at different bond dimensions. The upper (lower) two panels are data for three-chain (four-chain) systems. Extrapolations of the correlation lengths yield finite values. The slopes of $S$ versus $\ln{(\xi)}$ are approximately zero at large $\chi$.}
    \label{Gapped}
\end{figure}

Since we consider a periodic boundary condition in the transverse direction, the cylinder must have a gapped spectrum if we realize the CSL. We probe the gap indirectly by examining the correlation length and the central charge. For the three-chain cylinder, we obtain the correlation length at each bond dimension $\chi \in \{200,300, \cdots, 2000\}$ for both infinite MPSs $\psi_1$ and $\psi_s$ [Fig.~\ref{Gapped}(a)]. Extrapolations of the correlation length yield finite values as $\chi \rightarrow \infty$, which is compatible with a many-body gap in the system. We further confirm this conjecture by examining the scaling behavior of the entanglement entropy against $\ln{\left(\xi\right)}$ in Fig.~\ref{Gapped}(b). For a gapless one-dimensional system, its entanglement entropy scales as 
\begin{equation}
    S = \frac{c}{6} \ln{\left(\xi\right)} + S_0~,
\end{equation}
where $c \neq 0$ is the central charge, and $S_0$ is a nonuniversal constant~\cite{Calabrese_2009}. The entanglement entropy data in Fig.~\ref{Gapped}(b) bend flat as $\ln{\left(\xi\right)}$ increases, indicating the central charge is $0$. Similar to the $N=3$ case, the ground state for four-chain cylinders under the XC-4(0) boundary condition also has a finite correlation length and a zero central charge [Figs.~\ref{Gapped}(c) and \ref{Gapped}(d)].

\begin{figure}[tb]
\includegraphics[width=\columnwidth]{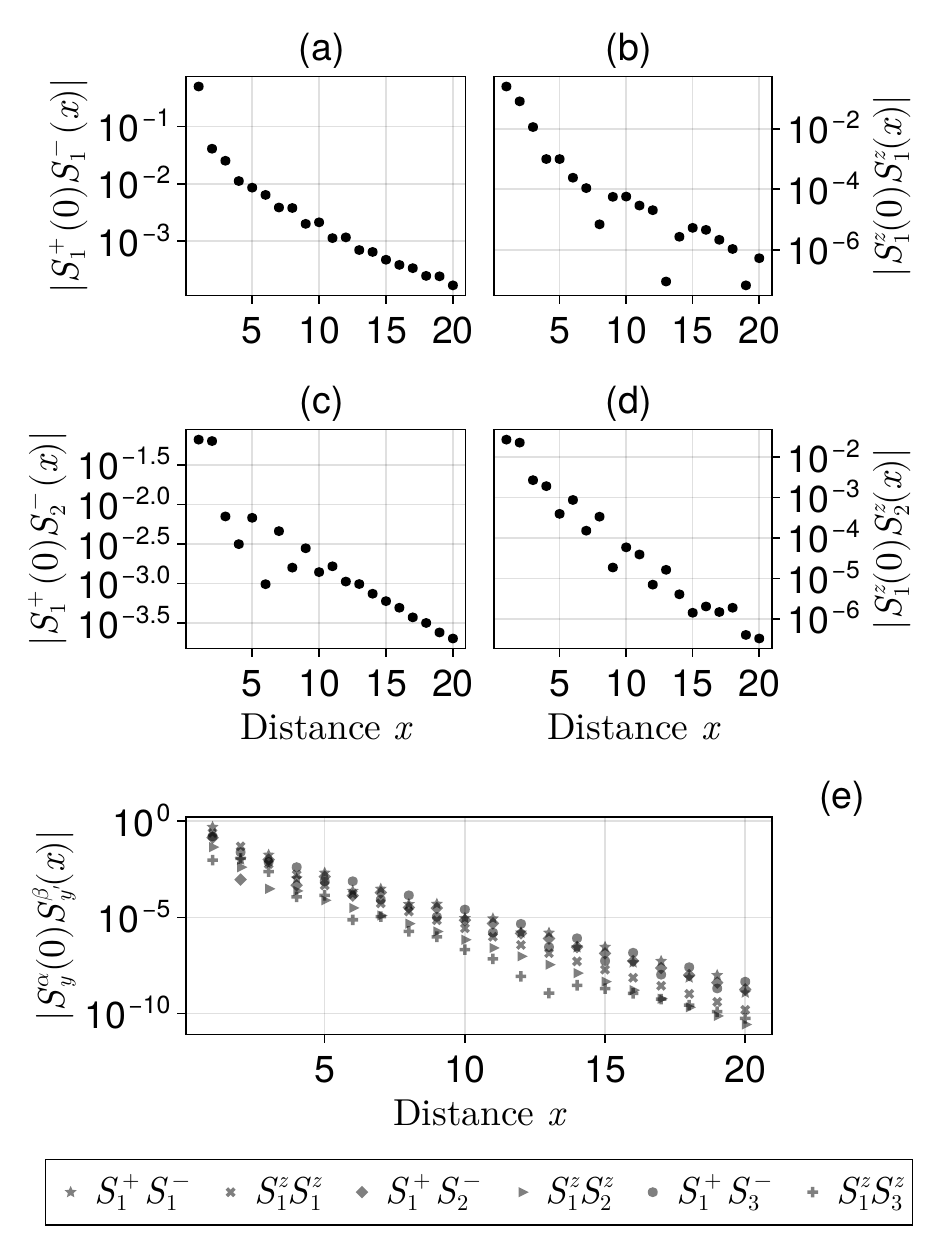}
\caption{Different two-point spin correlation functions for three-chain (a--d) and four-chain (e) systems. (a--d) The correlation functions are obtained from an infinite MPS $\psi_s$ lying in the identity sector with bond dimension $\chi = 2000$. Here, we only show functions involving the first and second chains. Other correlation functions with different chain indices can be easily obtained due to the translation symmetry in the transverse direction. Similarly, the correlation functions associated with the other ground state $\psi_1$ can be obtained by translation in the wire direction. (e) The correlation functions for the XC-4(0) cylinder. The wavefunction lies in the semion sector and has bond dimension $\chi = 1000$.} 
\label{fig.nomagnetism}
\end{figure}

We then use the states to examine spin operator correlation functions. The four small panels in Fig.~\ref{fig.nomagnetism} show correlation functions calculated by the wave function $\psi_s$ at $\chi = 2000$ for the three-chain cylinder. The correlation in the spin $XY$ plane shows perfect exponential decay, which is in agreement with the formation of descendants of the CSL. The other two panels have data affected more significantly by numerical errors. Still, the data follow modestly exponential decaying laws. The larger panel at the bottom shows the correlation functions for the XC-4(0) cylinder at $\chi = 1000$. All of the functions decay exponentially, which means the ground state does not have any magnetic order.

\subsection{Spontaneous symmetry breaking for $N=3$}

\begin{figure}
    \centering
    \includegraphics[width=\columnwidth]{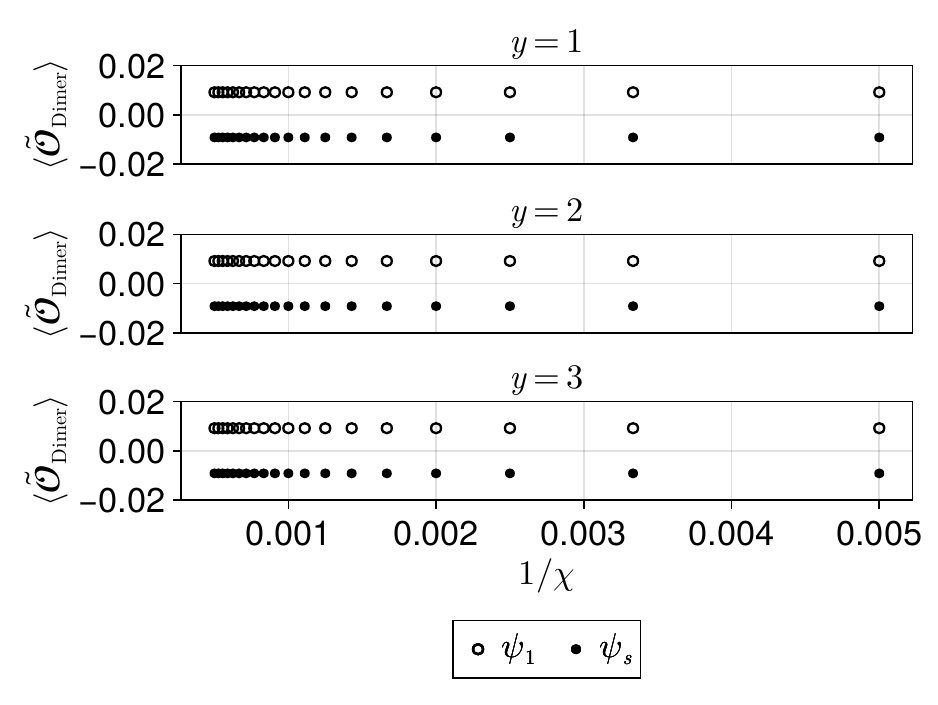}
    \caption{Values of $\expval{\Tilde{\cO}_{\text{Dimer}}}$ against the inverse of bond dimension for two degenerate ground states. We simulated the model at $\chi = 200,300,\cdots 2000$ and chose $x$ to be the horizontal coordinate of the site in the first unit cell. The quantities saturate at non-zero values, indicating a spontaneously broken translation symmetry.}
    \label{fig.XC3SSB}
\end{figure}

Based on Table~\ref{XC2GroundStateSummary}, three-chain cylinders should also break the translation symmetry as $N=5$ systems. We thus simulate the model with a six-site VUMPS ansatz to accommodate $S^z=0$ and permit translation-symmetry breaking. We have observed that the expectation value of $\mathcal{O}_{\text{Dimer}}$ defined in Eq.~\eqref{eqn.TOrderParam} exhibits relatively small oscillating behavior. Therefore, to single out the oscillating part, we consider
\begin{equation}
    \begin{split}
    \expval{\Tilde{\mathcal{O}}_\text{Dimer}(x,y)} &= \lim_{L\rightarrow \infty} \frac{1}{L} \sum_{x'} (-1)^{x' - x} \expval{\vec{S}_{x',y} \cdot \vec{S}_{x' + 1,y}}\\
    &= \frac{1}{2}\expval{\vec{S}_{x,y} \cdot \vec{S}_{x+1,y}} - \frac{1}{2}\expval{\vec{S}_{x+1,y} \cdot \vec{S}_{x+2,y}}.
    \end{split}
\end{equation}
The second equality holds since the wave function is an infinite uniform MPS, and it will serve as our way to obtain $\expval{\Tilde{\mathcal{O}}_\text{Dimer}}$. Figure \ref{fig.XC3SSB} shows this quantity for the two states $\psi_{1,s}$ at different bond dimensions. The quantity saturates at a nonzero value for both states, indicating a spontaneously broken translation symmetry.

\subsection{Entanglement spectra data for $N=3,6$}

\begin{figure}[tb]
\includegraphics[width=\columnwidth]{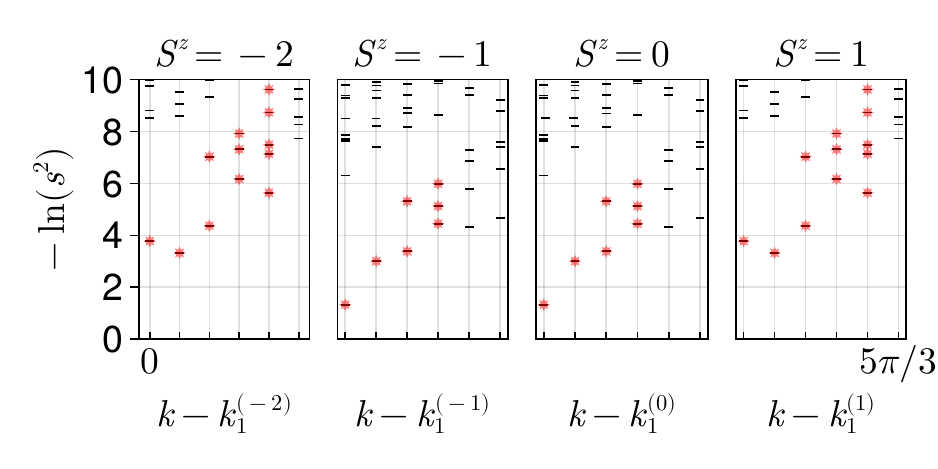}
\caption{Momentum-resolved spectra for the XC-6 cylinder with $32$ unit cells evaluated by $\psi_s$ at $\chi = 2000$. We label a few low-lying eigenvalues compatible with the level degeneracy $(1,1,2,3,5,7)$. Ticks on the horizontal axis for each panel are $0, \pi/3, \cdots, 5\pi/3$ from left to right.} 
\label{fig.SixChainES}
\end{figure}

The main text only contains entanglement spectra for four- and five-chain cylinders. We have also observed similar results for $N=6$ (see Fig.~\ref{fig.SixChainES}), yet the quality is not as superior as Figs.~\ref{FourChainES} and  \ref{FiveChainES}. In particular, we observed that the momentum of the lowest eigenvalue in the $S^z = -2$ and $S^z = 1$ sectors is wrong, which sits at the second allowed value for these two sectors. Furthermore, the degeneracy of each level becomes poor. Nevertheless, we suspect that increasing the bond dimension and potentially adding more wires may help improve the results.

\begin{figure}[tb]
    \centering
    \includegraphics[width=\linewidth]{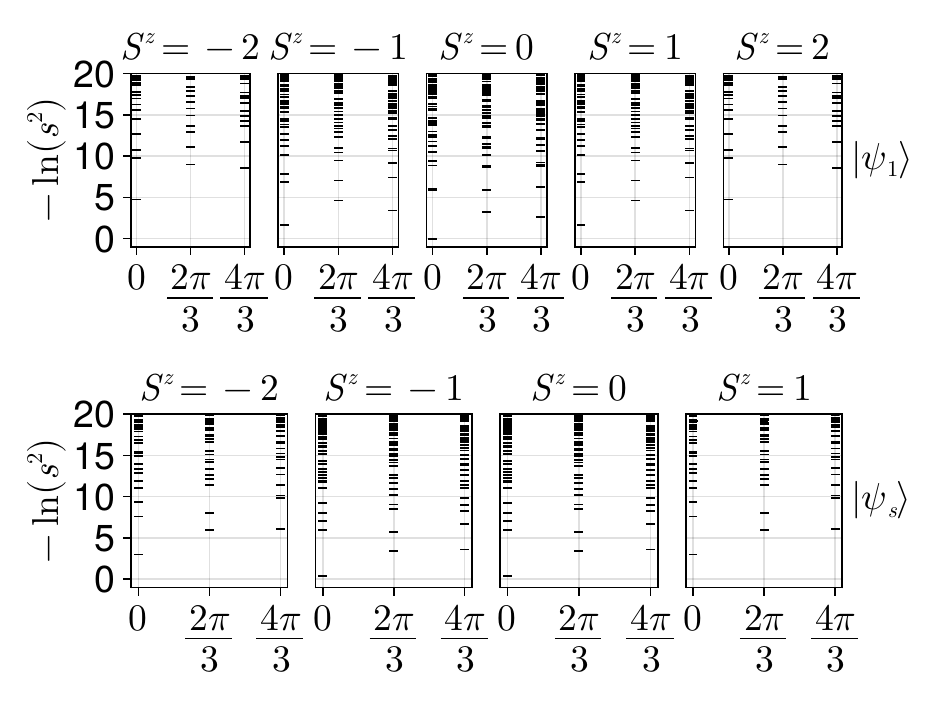}
    \caption{Momentum-resolved entanglement spectra of infinite MPSs at $\chi = 2000$ for the XC-3 cylinder. We similarly shift the spectra as described in the caption of Fig.~\ref{FiveChainES}. The upper five figures describe state $\psi_1$ in the identity sector, while the state $\psi_s$ yields the lower four figures.}
    \label{fig.ThreeChainES}
\end{figure}

We also obtain the entanglement spectra for both $\psi_1$ and $\psi_s$ associated with the three-chain cylinder (Fig.~\ref{fig.ThreeChainES}). Due to the small circumference, one cannot unambiguously identify the chirality or degeneracy pattern. Still, the spectra are consistent with the theoretical expectations.

\subsection{Results at a different bond dimension for $N=5,6$}

The data for five- and six-chain systems in the main text come solely from MPSs at bond dimension $\chi = 2000$. In contrast, we obtained a series of infinite MPSs during VUMPS simulations for $N=3,4$ and confirmed convergence by looking at correlation lengths and entanglement entropy. To support that the results for $N=5,6$ are already faithful, we include measurements of the same quantities as shown in the main text using MPSs at $\chi = 1000$.

\begin{figure}[tb]
    \centering
    \includegraphics[width=\linewidth]{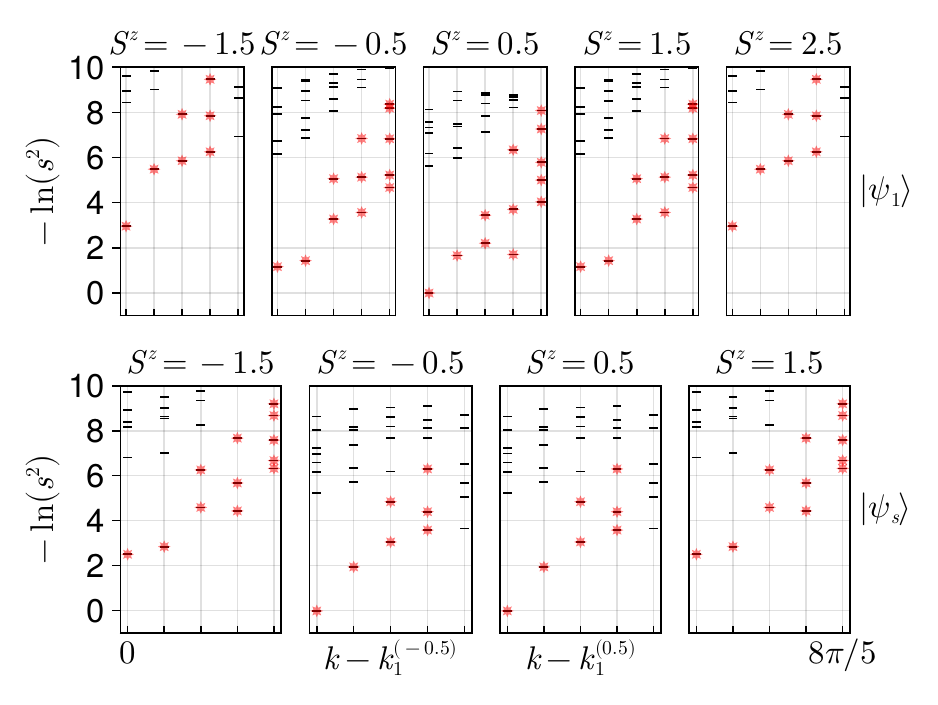}
    \caption{Momentum-resolved entanglement spectra of MPSs at $\chi = 1000$ for a five-chain finite cylinder with $31$ unit cells similar to Fig.~\ref{FiveChainES}. The spectra have almost an identical low-lying part compared with Fig.~\ref{FiveChainES}.}
    \label{fig.FiveChainES1000}
\end{figure}

First, we examine the entanglement spectra for $\psi_1$ and $\psi_s$ in a five-chain systems at bond dimension $\chi = 1000$ (Fig.~\ref{fig.FiveChainES1000}). The low-lying states exhibit the expected pattern of a Laughlin state. Noticeable differences with Fig.~\ref{FiveChainES} at $\chi = 2000$ occur at higher levels in the spectra. In particular, the $\chi = 2000$ spectra have eigenvalues associated with definite transverse momenta at entanglement energy $-\ln{(s^2)} \gtrsim 10$. These eigenvalues get mixed with each other when we resolve $k_y$ at $\chi = 1000$. Thus, we exclude them from Fig.~\ref{fig.FiveChainES1000}.

\begin{figure}[tb]
    \centering
    \includegraphics[width=\linewidth]{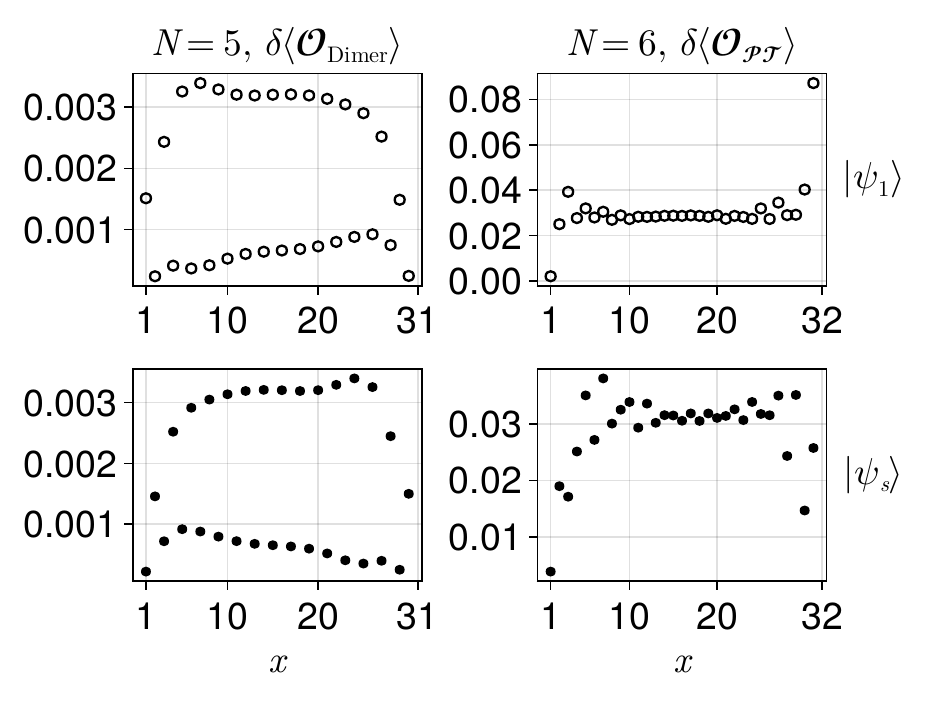}
    \caption{Relative differences of the order parameters defined in Eqs.~\eqref{eqn.TOrderParam} and \eqref{eqn.PTOrder} evaluated by MPSs at $\chi = 1000$ and $\chi = 2000$. The first (second) column has data for the five-chain (six-chain) system. The two figures in the first (second) row correspond to the state in the identity (semion) sector. The relative difference is defined as $\delta f = |f_{1000} - f_{2000}|/|f_{2000}|$ where $f_\chi$ is the result at bond dimension $\chi$.}
    \label{fig.SSB1000}
\end{figure}

Second, we evaluate the order parameters for spontaneous symmetry breakings using states at $\chi = 1000$. The data also exhibit the same symmetry-breaking behaviors. In Fig.~\ref{fig.SSB1000}, we show the relative difference between the two sets of data: one is at bond dimension $\chi = 1000$, and the other one corresponds to Fig.~\ref{fig.TSSB}. The maximal relative difference of $\expval{\mathcal{O}_{\text{Dimer}}}$ near the system's center is around $0.3\%$, and the same maximal relative error for $\expval{\mathcal{O}_{\mathcal{P}\mathcal{T}}}$ is around $3\%$.

\begin{figure}[tb]
    \centering
    \includegraphics[width=\linewidth]{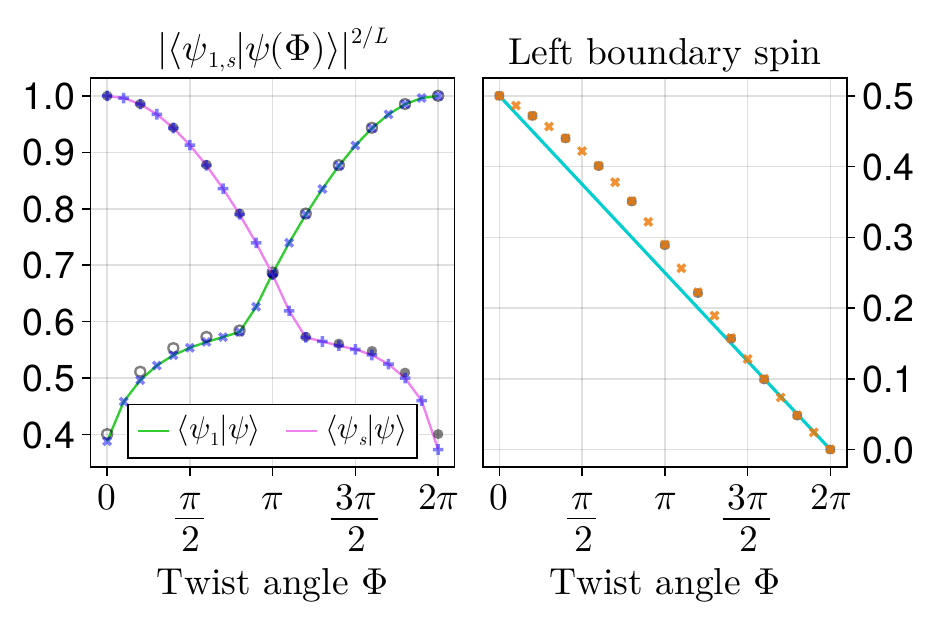}
    \caption{Spin pumping simulations for the XC-6(0) cylinder at $\chi = 1000$. The results overlay on the top of the data in Fig.~\ref{fig.SpinPumping}. In both panels, the old data (circles) are in gray. The reference states $\psi_1$ and $\psi_s$ in the second panel are the same wave functions considered in Fig.~\ref{fig.SpinPumping}, whose bond dimension $\chi$ is $2000$.}
    \label{fig.XC60Spin1000}
\end{figure}

Finally, we perform the flux insertion at $\chi = 1000$ for the XC-6(0) cylinder (Fig.~\ref{fig.XC60Spin1000}). As anticipated, the spin transferred from one edge to the other is still precisely $1/2$. At the same time, the MPS $\psi(\Phi)$ evolves from the semion sector to the identity sector. We note that the difference between the two sets of data ($\chi = 1000,2000$) in the first panel of Fig.~\ref{fig.XC60Spin1000} is small, while the difference in the second panel is negligible.

\bibliography{reference, data}

\end{document}